\documentclass[referee]{aa} 

\usepackage[dvips]{graphicx}
\usepackage{epsfig}
\usepackage{txfonts}

\usepackage{natbib}
\bibpunct{(}{)}{;}{a}{}{,}

\newcommand{\rot}{\vec{\rm rot}\,}
\renewcommand{\div}{{\rm div}\,}

\begin{document}

\title{Relativistic stabilisation of the diocotron instability in a
  pulsar ``cylindrical'' electrosphere.}

\author{J\'er\^ome P\'etri \inst{1}}

\offprints{J. P\'etri}

\institute{Max-Planck-Institut f\"ur Kernphysik, Saupfercheckweg 1,
  69117 Heidelberg, Germany.}

\date{Received / Accepted}

\titlerunning{Relativistic stabilisation of the diocotron instability
  in a pulsar electrosphere}

\authorrunning{P\'etri}

\abstract
{The physics of the pulsar inner magnetosphere remains poorly
  constrained by observations. Although about 2000~pulsars have been
  discovered to date, only little is known about their emission
  mechanism.  Large vacuum gaps exist in the magnetosphere and a
  non-neutral plasma fills partially the neutron star surroundings to
  form an electrosphere.}
{In a previous work, we showed that the differentially rotating
  equatorial disk in the pulsar's electrosphere is diocotron unstable
  in the non-relativistic regime. In this paper, we extend these
  results and study the relativistic and electromagnetic stabilisation
  effects by including the magnetic field perturbation and allow for
  relativistic speeds of the guiding centre, in a self-consistent
  manner. We use the electric drift approximation, valid for
  low-density plasmas.}
{We linearise the coupled relativistic cold-fluid and Maxwell
  equations in the electric drift approximation. The non-linear
  eigenvalue problem for the perturbed azimuthal electric field is
  solved numerically with standard technics for boundary value
  problems like the shooting method. The spectrum of the relativistic
  diocotron instability in a non-neutral plasma column confined
  between two cylindrically conducting walls is computed.}
{For low-speed motions, we recover the eigenfunctions and eigenspectra
  of the non-relativistic diocotron instability. Our algorithm is also
  checked in the relativistic planar diode geometry for which an
  analytical expression of the dispersion relation is known. As
  expected, when the relativistic and electromagnetic effects become
  significant, the diocotron instability tends to stabilise. In
  cylindrical geometry, for some special rotation profile, all
  azimuthal modes~$l$ are completely suppressed for sufficiently
  relativistic flows. However, for the profile relevant to the
  electrosphere, depending on the exact rotation curves, the growth
  rates can either significantly decrease till they vanish or persist
  for moderate~$l$.}
{The non-neutral plasma flowing in the pulsar electrosphere approaches
  the speed of light when reaching the light-cylinder. Therefore,
  relativistic and electromagnetic effects are important. They are
  capable to completely suppress the diocotron instability.
  Nevertheless, results are sensitive to the tail of the rotation
  curves. Therefore, particle diffusion across the magnetic field due
  to the diocotron instability only works efficiently close to the
  neutron star surface. }
   

\maketitle

\section{INTRODUCTION}

The detailed structure of charge distribution and electric-current
circulation in the closed magnetosphere of a pulsar remains poorly
understood. Although it is often assumed that the plasma fills the
space entirely and corotates with the neutron star, it is on the
contrary very likely that it only partly fills it, leaving large
vacuum gaps between plasma-filled regions. The existence of such gaps
in aligned rotators has been very clearly established by
\cite{1985MNRAS.213P..43K, 1985A&A...144...72K}.  Since then, a number
of different numerical approaches to the problem have confirmed their
conclusions, including some work by \cite{1989Ap&SS.158..297R},
\cite{1989Ap&SS.161..187S}, \cite{1993A&A...268..705Z},
\cite{1993A&A...274..319N}, \cite{1994ApJ...431..718T},
\cite{2002ASPC..271...81S}, and by ourselves
\citep{2002A&A...384..414P}.  This conclusion about the existence of
vacuum gaps has been reached from a self-consistent solution of
Maxwell's equations in the case of the aligned rotator.  Moreover,
\cite{2001MNRAS.322..209S} have shown by numerical modelling that an
initially filled magnetosphere like the Goldreich-Julian model evolves
by opening up large gaps and stabilises to the partially filled and
partially void solution found by \cite{1985MNRAS.213P..43K}, and also
by \cite{2002A&A...384..414P}.  The status of models of the pulsar
magnetospheres, or electrospheres, has recently been critically
reviewed by \cite{2005RMxAC..23...27M}.  A solution with vacuum gaps
has the peculiar property that those parts of the magnetosphere that
are separated from the star's surface by a vacuum region are not
corotating and so suffer differential rotation.
  
This raises the question of the stability of such a charged plasma
flow.  The differential rotation in the equatorial, non neutral disk
induces the so-called diocotron and magnetron instabilities that are
well known to plasma physicists \citep{1980PhFl...23.2216O,
  Davidson1990, 1992PhFlB...4.2720O}.  In the inner parts of the
magnetosphere, far from the light cylinder, the instability reduces to
its electrostatic form, the diocotron instability. The linear
development of the diocotron instability of a thin differentially
rotating charged disk was studied by \cite{2002A&A...387..520P} and
\cite{2007A&A...464..135P} and shown to proceed at a growth rate
comparable to the star's rotation rate.  The non linear development of
this instability was studied by \cite{2003A&A...411..203P}, in the
framework of an infinitely thin disk model. They have shown that the
instability causes a cross-field transport of these charges in the
equatorial disk, evolving into a net out-flowing flux of charges.
\cite{2002ASPC..271...81S} have numerically studied the problem, and
concluded that this charge transport tends to fill the gaps with
plasma. The appearance of a cross-field electric current as a result
of the diocotron instability has been observed by
\cite{2002AIPC..606..453P} in laboratory experiments in which charged
particles were continuously injected in the plasma column trapped in a
Malmberg-Penning configuration.
    
The aim of this work is to extend the previous work done by
\cite{2007A&A...464..135P} on the diocotron instability by including
electromagnetic and relativistic effects. Indeed, when the plasma
approaches the light-cylinder, the guiding centre motion becomes
relativistic and magnetic perturbations become significant.

The relativistic aspect of the diocotron instability have already been
investigated in the planar diode geometry by
\cite{1987PhLA..125...61D, 1988PhFl...31.1727D}.  They clearly
demonstrated the stabilisation due to electromagnetic effects.
Applications to the planar magnetron geometry was investigated by
\cite{1992PhFlB...4.3396A}.

In this paper we present a numerical analysis of the linear growth
rate of the relativistic diocotron instability for a plasma column.
The paper is organised as follows.  In Sect.~\ref{sec:Setup}, we
describe the initial setup of the plasma column consisting of an
axially symmetric equilibrium between two conducting walls. We give
several equilibrium profiles useful for the study of the diocotron
instability in different configurations. In Sect.~\ref{sec:AnalLin},
the non-linear eigenvalue problem satisfied by the perturbed azimuthal
electric field component is derived. The algorithm to solve the
eigenvalue problem is checked against known analytical results in the
non-relativistic cylindrical limit as well as in the relativistic
planar diode geometry, Sect.~\ref{sec:Check}. Then, applications to
some typical equilibrium configuration are shown in
Sect.~\ref{sec:Results}. First we consider a plasma column with
constant diocotron frequency. Next, we study the effect of the
cylindrical geometry (curvature of the flow) and the transition to the
planar diode limit. Finally, the stabilisation of the diocotron
instability in a pulsar electrosphere is demonstrated when the flow
becomes ultra-relativistic.  The conclusions and the possible
generalisation are presented in Sect.~\ref{sec:Conclusion}.

\section{INITIAL SETUP}
\label{sec:Setup}

We study the motion of a non neutral plasma column of infinite axial
extend along the $z$-axis, in the electric drift approximation.  This
approximation is well-suited for low-density plasmas satisfying
$\omega_{\rm p}^2 \ll \omega_{\rm c}^2$ where $\omega_{\rm p}$ is the
plasma frequency and $\omega_{\rm c}$ the cyclotron frequency.  The
geometric configuration is the same as in \cite{2007A&A...464..135P}.
However, perturbations in magnetic field induced by the relativistic
motion of the charges are now taken into account in a fully
self-consistent manner. Therefore, we have to solve the full set of
Maxwell equations.

In this section, we briefly summarise the equilibrium conditions
imposed on the plasma and give some typical examples of equilibrium
configurations for specified velocity, density and electric field
profiles.

\subsection{Equilibrium of the plasma column}

We consider a single-species non-neutral plasma consisting of
particles with mass~$m_{\rm e}$ and charge~$q$ trapped between two
cylindrically conducting walls located at $r = W_1$ and $r = W_2 >
W_1$. The plasma column itself is confined between $R_1 \ge W_1$ and
$R_2 \le W_2$.  This allows us to take into account vacuum regions
between the plasma and the conducting walls. We adopt cylindrical
coordinates denoted by~$(r,\varphi,z)$ and the corresponding basis
vectors~$(\vec{e}_{\rm r},\vec{e}_{\rm \varphi},\vec{e}_{\rm z})$.  

In the equilibrium configuration, the particle number density
is~$n_{\rm e}(r)$ and the associated charge density is~$\rho_{\rm
  e}(r) = q \, n_{\rm e}(r)$.  Particles evolve in a cross electric
and magnetic field such that the equilibrium magnetic field is
directed along the~$z$-axis whereas the equilibrium electric field is
directed along the $r$-axis. The electric field induced by the plasma
itself, $\vec{E}_\mathrm{p}$, is
\begin{eqnarray}
  \label{eq:ErTot}
  \vec{E}_{\rm p} & = & E_{\rm r} \, \vec{e}_{\rm r} .
\end{eqnarray}
The magnetic field is made of two parts, the first is an imposed
external applied field, $\vec{B}_0$, assumed to be uniform in the
region outside the plasma column, and the second a plasma induced
field, $\vec{B}_\mathrm{p}$
\begin{equation}
  \label{eq:ETot}
  \vec{B} = \vec{B}_\mathrm{p} + \vec{B}_0 = B_{\rm z} \, \vec{e}_{\rm z} .
\end{equation}
Therefore, for azimuthally symmetric equilibria, the steady-state
Maxwell-Gauss and Maxwell-Amp\`ere equations satisfy
\begin{eqnarray}
  \label{eq:PoissonDiocRelat}
  \frac{1}{r} \, \frac{\partial}{\partial r} ( r \, E_{\rm r}) & = &
  \frac{\rho_{\rm e}}{\varepsilon_0} \\
  \label{eq:AmpereDiocRelat}
  \frac{\partial B_{\rm z}}{\partial r} & = & - \mu_0 \, \rho_{\rm e} \, v_\varphi .
\end{eqnarray}
In the electric drift approximation, particle trajectories are
described by the guiding centre motion so that the flow velocity is
only azimuthal and expressed as
\begin{equation}
  \label{eq:vDiocRelat}
  \vec{v} = v_\varphi \, \vec{e}_\varphi
  = - \frac{E_{\rm r}}{B_{\rm z}} \, \vec{e}_\varphi .
\end{equation}
In order to avoid unphysical superluminal motion of the guiding
centre, $v<c$, this model can only be applied for $E_{\rm r} < c \,
B_{\rm z}$. In the opposite case, i.e. when $E_{\rm r} > c \, B_{\rm
  z}$, inertia effects become important and should be included in the
equation of motion. The most general instability in a non-neutral
plasma taking into account inertia and relativistic effects is the
so-called magnetron instability. The full self-consistent response of
the plasma to this instability will be studied in a forthcoming paper
(removing the electric drift approximation, replaced by the
relativistic momentum equation). In the case of pulsar magnetospheres,
let us give an estimate of the distance to which this approximation
remains valid. The drift motion, Eq.~(\ref{eq:vDiocRelat}) applies
roughly when the Larmor radius of the particles~$R_{\rm g}$ is less
than the radius of their orbit~$R$.  Assuming relativistic speeds,
with a guiding centre motion at the star rotation speed~$\Omega_*$, we
get
\begin{equation}
  \label{eq:Larmor}
  R_{\rm g} = \frac{\gamma \, m_{\rm e} \, c}{|q| \, B} =
  \frac{m_{\rm e} \, c}{|q| \, B \, \sqrt{1 - R^2 \, \Omega_*^2 / c^2}} < R
\end{equation}
Let $R_*$ and $B_*$ be the radius of the neutron star and the
surface magnetic field respectively. For a dipolar vacuum magnetic
field, we have
\begin{equation}
  \label{eq:BPulsar}
  B(R) = B_* \, \frac{R_*^3}{R^3}
\end{equation}
Condition Eq.~(\ref{eq:Larmor}) can be cast into
\begin{equation}
  \label{eq:Conditions}
  \frac{R^2}{\sqrt{1-R^2\,\Omega_*^2/c^2}} <
  \frac{|q| \, B_* \, R_*^3}{m_{\rm e} \, c} \equiv Y^2
\end{equation}
We introduced the parameter
\begin{equation}
  \label{eq:Yparam}
  Y = \sqrt{\frac{|q| \, B_* \, R_*^3}{m_{\rm e} \, c}}
\end{equation}
Thus, we find that a rough estimate of the largest radius where the
drift approximation remains valid is
\begin{equation}
  \label{eq:RayonMax}
  \frac{R_{\rm max}}{R_{\rm L}} \lesssim \frac{Y^2}{\sqrt{2} \, R_{\rm L}^2}
  \, \sqrt{ \sqrt{1 + 4 \, \frac{R_{\rm L}^4}{Y^4} } - 1 }
\end{equation}
Applying to pulsar with period of the order one second, $P=1$~s and
magnetic fields of the order~$B_*=10^8$~T, assuming a typical
neutron star radius of~$R_*=10$~km, we get
\begin{equation}
  \label{eq:RL}
  R_{\rm L} = 4771 \, R_* \;\;\; ; \;\;\; Y = 5076 \, R_{\rm L} \gg R_{\rm L}
\end{equation}
Thus $R_{\rm max} \lesssim R_{\rm L}$. The approximation falls down
only very close to the light cylinder~$R_{\rm L}$.  For millisecond
pulsars, $P=1$~ms and $B_*=10^5$~T, the same conclusion applies
because
\begin{equation}
  R_{\rm L} = 4.771 \, R_* \;\;\; ; \;\;\; Y = 160531 \, R_{\rm L} \gg R_{\rm L}
\end{equation} 
We conclude that the electric drift approximation remains valid to
good accuracy for electrospheric plasmas confined within the light
cylinder.

For a constant density profile in the plasma column with $W_1=0$, the
electric drift approximation, Eq.~(\ref{eq:vDiocRelat}), corresponds
to a circular motion at the diocotron frequency defined by, (see for
instance \cite{Davidson1990}),
\begin{equation}
  \label{eq:DiocFreq}
  \omega_{\rm D} = \frac{\omega_{\rm p}^2}{2\,\omega_{\rm c}}
\end{equation}
The relativistic plasma and cyclotron frequencies are given
respectively by
\begin{eqnarray}
  \label{eq:omega_p}
  \omega_{\rm p}^2 & = & \frac{\rho_{\rm e} \, q}{\gamma \, m_e \, \varepsilon_0} \\
  \label{eq:omega_c}
  \omega_{\rm c} & = & \frac{q \, B_{\rm z}}{\gamma \, m_e}
\end{eqnarray}
where $\gamma = 1 / \sqrt{1 - v_\varphi^2 / c^2}$ corresponds to the
bulk Lorentz factor of the flow.

We assume that the electric field induced by the plasma vanishes at
the inner wall, at $r=W_1$, i.e.
\begin{equation}
  \label{eq:ErCL}
  \vec{E}_\mathrm{p}(W_1) = \vec{0} .  
\end{equation}
Integrating Eq.~(\ref{eq:PoissonDiocRelat}) therefore gives for the
electric field generated by the plasma,
\begin{equation}
  \label{eq:Ep}
  \vec{E}_\mathrm{p}(r) = \frac{1}{\varepsilon_0\,r} \, \int_{W_1}^r 
  \rho_{\rm e}(r') \, r' \, dr' \, \vec{e}_{\rm r}
\end{equation}
For the magnetic field induced by the plasma, we solve
Eq.~(\ref{eq:AmpereDiocRelat}) with the boundary condition $B_{\rm
  z}(R_2) = B_0$. This simply states that the total magnetic field
outside the plasma column has to match the magnetic field imposed by
an external device.

Any equilibrium state is completely determined by the following four
quantities, the total radial electric field, $E_{\rm r}$, the total
axial magnetic field, $B_{\rm z}$, the charge density, $\rho_{\rm e}$,
and the azimuthal speed of the guiding centre, $v_\varphi$.
Prescribing one of these profiles, the remaining three are found
self-consistently by solving the set of
Eqs.~(\ref{eq:PoissonDiocRelat}), (\ref{eq:AmpereDiocRelat}) and
(\ref{eq:vDiocRelat}). We give some typical examples in the next
sections in which the velocity profile, the density profile or the
electric field is imposed.

\subsection{Specified velocity profile}

Let us first assume that the velocity profile~$v_\varphi = r \,
\Omega$ is prescribed.  This case is well-suited for the study of the
pulsar's electrosphere in which the plasma is in differential
rotation.  Although the magnetic field is probably nearly dipolar in a
pulsar, we assume a uniform magnetic field in vacuum in order to
remain fully self-consistent. We know from the previous analysis,
\citep{2007A&A...464..135P}, that the growth rates are not very
sensitive to the magnetic field structure. As already noticed in this
work, the differential rotation is essential to the presence of the
diocotron instability.  The other equilibrium quantities, $(E_{\rm r},
B_{\rm z}, \rho_{\rm e})$, are easily derived from $\Omega$.  Indeed,
inserting $\rho_{\rm e}$ from Eq.~(\ref{eq:PoissonDiocRelat}) and
$E_r$ from Eq.~(\ref{eq:vDiocRelat}) into Maxwell-Amp\`ere equation
(\ref{eq:AmpereDiocRelat}), the magnetic field satisfies a first order
ordinary linear differential equation
\begin{eqnarray}
  \label{eq:BzEDO}
  \frac{\partial B_{\rm z}}{\partial r} & = & \frac{\gamma^2 \, \beta}{r} \, B_{\rm z} \, 
  \frac{\partial}{\partial r} ( r \, \beta ) = 
  \frac{\gamma^2 \, \Omega}{c^2} \, B_{\rm z} \, \frac{\partial}{\partial r} 
  \left( r^2 \, \Omega \right) .
\end{eqnarray}
The Lorentz factor of the flow is
\begin{eqnarray}
  \label{eq:LorentzFlow}
  \gamma & = & \frac{1}{\sqrt{ 1 - \beta^2 }} \\
  \beta & = & \frac{r \, \Omega}{c}
\end{eqnarray}
From Poisson equation, Eq.~(\ref{eq:PoissonDiocRelat}), and the
definition of the plasma frequencies, Eqs.~(\ref{eq:omega_p}) and
(\ref{eq:omega_c}), the charge density is found by
\begin{eqnarray}
  \label{eq:rhosBz}
  \frac{\rho_{\rm e}}{\varepsilon_0 \, B_{\rm z}} & = & 
  \frac{\omega_{\rm p}^2}{\omega_{\rm c}} =
  - \frac{\gamma^2}{r} \, \frac{\partial}{\partial r} \left( r^2 \, \Omega \right) =
  - \gamma^2 \, \left[ 2 \, \Omega + r \, \frac{\partial\Omega}{\partial r} \right]
\end{eqnarray}
The electric field is recovered from Poisson equation,
Eq.~(\ref{eq:PoissonDiocRelat}), or, according to the electric drift
approximation, by Eq.~(\ref{eq:vDiocRelat}) leading to the charge
density
\begin{equation}
  \label{eq:DensiteCharge}
  \rho_{\rm e} = - \varepsilon_0 \, \left[ \frac{B_{\rm z}}{r} \,
    \frac{\partial}{\partial r} \left( r^2 \, \Omega \right)
    + r \, \Omega \, \frac{\partial B_{\rm z}}{\partial r} \right]
\end{equation}
consistent with Eqs.~(\ref{eq:BzEDO}) and (\ref{eq:rhosBz}).

\subsection{Specified density profile}
\label{sec:ProfDens}

A simple and useful charge density profile to study the relativistic
effect on the diocotron instability is given by a constant diocotron
frequency in the whole plasma column, Eq.~(\ref{eq:DiocFreq}), such
that
\begin{equation}
  \label{eq:DiocRect}
  \omega_{\rm D}(r) = \left \lbrace
    \begin{array}{lcl}
      0   & , & W_1 \le r \le R_1 \\
      \omega_{\rm D} = {\rm const} & , & R_1 \le r \le R_2 \\
      0   & , & R_2 \le r \le W_2
    \end{array}
  \right.
\end{equation}
Using Eq.~(\ref{eq:rhosBz}), the rotation profile is deduced by
integrating an ordinary differential equation for $\Omega$
\begin{equation}
  \label{eq:EDOOmega}
  \frac{\partial\Omega}{\partial r} = - \frac{2}{r} \, \left[ \Omega 
    + \omega_{\rm D} \, \left( 1 - \frac{r^2 \, \Omega^2}{c^2} \right) \right]
\end{equation}
with the boundary condition that $\Omega$ vanishes at the inner plasma
edge, $\Omega(R_1) = 0$, to insure consistency with
Eq.~(\ref{eq:ErCL}).  Knowing $\Omega$, the same procedure as in the
previous subsection for a specified velocity profile is applied, i.e.
the magnetic field is calculated according to Eq.~(\ref{eq:BzEDO}).

\subsection{Specified electric field}
\label{sec:Electric}

It is also possible to specify the equilibrium radial electric field.
An interesting case is given by
\begin{equation}
  \label{eq:Er0}
  E_{\rm r}(r) = \left \lbrace
    \begin{array}{lcl}
      0  & , & W_1 \le r \le R_1 \\
      - B_0 \, \displaystyle{ \frac{\sinh \alpha \, ( r - R_1 )}
      {\cosh \alpha \, ( R_2 - R_1 )} } & , & R_1 \le r \le R_2  \\
      - B_0 \, \displaystyle{ \frac{R_2 \, \tanh \alpha \, ( R_2 - R_1 )}{r} }
      & , & R_2 \le r \le W_2
    \end{array}
  \right.
\end{equation}
$\alpha$ is a constant useful to adjust the maximal speed of the
column at $R_2$
\begin{equation}
  \label{eq:Vmax}
  v_{\rm max} = \tanh \alpha \, ( R_2 - R_1 ).
\end{equation}
The equilibrium electric profile, Eq.~(\ref{eq:Er0}), enables us to
investigate the influence of the cylindrical geometry compared to the
planar diode geometry. Indeed, in the limit of small curvature of the
column, i.e. when $R_2-R_1\ll R_1$, the eigenvalue problem in
cylindrical geometry reduces to the planar diode case.  The charge
density is found by Eq.~(\ref{eq:PoissonDiocRelat}) whereas the
magnetic field is solution of an ordinary differential equation
\begin{equation}
  \label{eq:BzEDOEr}
  \frac{\partial B_{\rm z}^2}{\partial r} = 
  \frac{1}{r^2 \, c^2} \, \frac{\partial}{\partial r} 
\left( r^2 \, E_{\rm r}^2 \right)
\end{equation}
Finally, the velocity is given by the electric drift approximation,
Eq.~(\ref{eq:vDiocRelat}).

\section{LINEAR ANALYSIS}
\label{sec:AnalLin}

In this section, we show how to derive the eigenvalue problem for the
diocotron instability in the relativistic regime.  Let us start with
the full set of non-linear equations governing the plasma.

\subsection{Equations of motion}

Each particle evolves in the self-consistent electromagnetic field
partly imposed by an external device and partly induced by the plasma
itself. The motion of the plasma column is governed by the
conservation of charge, the electric drift approximation, and the full
set of Maxwell equations, respectively,
\begin{eqnarray}
  \label{eq:Continuite}
  \frac{\partial \rho_{\rm e}}{\partial t} + 
  \div (\rho_{\rm e} \, \vec{v}) & = & 0 \\
  \label{eq:Vit}
  \vec{v} & = & \frac{\vec{E} \wedge \vec{B}}{B^2} \\
  \label{eq:MaxFar}
  \rot \vec{E} & = & - \frac{\partial \vec{B}}{\partial t} \\
  \label{eq:MaxAmp}
  \rot \vec{B} & = & \mu_0 \, \vec{j} +
  \varepsilon_0 \, \mu_0 \, \frac{\partial \vec{E}}{\partial t} \\
  \label{eq:MaxGauss}
  \div \vec{E} & = & \frac{\rho_{\rm e}}{\varepsilon_0} \\
  \label{eq:MaxB}
  \div \vec{B} & = & 0
\end{eqnarray}
For a non-neutral plasma, the current density is related to the charge
density by
\begin{equation}
  \label{eq:DensiteCourant}
  \vec{j} = \rho_{\rm e} \, \vec{v}
\end{equation}
We apply the standard linear perturbation theory.  All perturbations
of physical quantities~$X$ like electric field, density, and velocity
components, are expressed by the expansion
\begin{equation}
  \label{eq:Expa}
  X(r,\varphi,t) = X(r) \, e^{i \, (l\,\varphi - \omega\,t)}
\end{equation}
where $l$ is the azimuthal mode and $\omega$ the eigenfrequency.
Therefore, we do not take into account the finite thickness of the
disk. The whole cylinder moves as one block in the vertical direction.
However, because the disk of the pulsar electrosphere has a finite
vertical dimension, denoted by~$H$, these modes should also depend on
the vertical coordinate~$z$. It is possible to extend this cylindrical
model to modes depending on~$z$ by introducing the following expansion
\begin{equation}
  \label{eq:ExpaRZ}
  X(r,\varphi,z,t) = X(r) \, e^{i \, ( k \, z + l \, \varphi - \omega \, t)} .
\end{equation}
The technic is similar to the one used to investigate wave propagation
in accretion disks of finite thickness. Several methods could be
applied. For instance, \cite{1991ApJ...378..656N} used a series
expansion in $z/H$ of the Lagrangian displacement of a fluid element
in order to study trapped adiabatic oscillations in accretion disks.
Looking for solutions that are separable in the radial and vertical
direction is another mean to seek for the perturbations. This was done
for instance by \cite{1987PASJ...39..457O} for trapped oscillations in
relativistic accretion disks and by \cite{1989PASJ...41..745K} for
low-frequency corrugation waves in an isothermal relativistic
accretion disk. To get insight in the properties of the oscillations,
\cite{1997ApJ...476..589P} performed a radial and vertical WKB
analysis.
  
In our case, for a small vertical extension of the disk, $k\,H\gg1$, a
series expansion in $z/H$ would be appropriate. Indeed, regions with
$z\gg H$ will not contribute significantly to the electromagnetic
field because of the oscillating term~$e^{i\,k\,z}$.  However, this
more general study is left for future work.

\subsection{Linearisation}

We study the stability of the plasma column around the equilibrium
mentioned in the previous section.  An expansion to first order for
the electromagnetic field around the equilibrium $(\vec{E}^0, B_{\rm
  z}^0)$ leads us to
\begin{eqnarray}
  \label{eq:DE}
  \vec{E} & = & \vec{E}^0 + \delta \vec{E} \\
  \label{eq:DBz}
  B_{\rm z} & = & B_{\rm z}^0 + \delta B_{\rm z}   
\end{eqnarray}
and the same for the charge and current density
\begin{eqnarray}
  \label{eq:Dj}
  \vec{j} & = & \vec{j}^0 + \delta \vec{j} \\
  \label{eq:Drho}
  \rho_{\rm e} & = & \rho_{\rm e}^0 + \delta \rho_{\rm e}   
\end{eqnarray}
Linearising the set of Maxwell equations, we have
\begin{eqnarray}
  \frac{1}{r} \, \frac{\partial}{\partial r} ( r \, \delta \, E_{\rm r}) +
  i \, \frac{l}{r} \, \delta E_\varphi & = & 
  \frac{\delta \rho_{\rm e}}{\varepsilon_0} \\
  \frac{1}{r} \, \frac{\partial}{\partial r} ( r \, \delta \, E_\varphi) -
  i \, \frac{l}{r} \, \delta E_{\rm r} & = & i \, \omega \, \delta B_{\rm z} \\
  i \, \frac{l}{r} \, \delta B_{\rm z} & = & \mu_0 \, \delta j_{\rm r} - 
  i \, \frac{\omega}{c^2} \, \delta E_{\rm r} \\
  - \frac{\partial}{\partial r} \delta B_{\rm z} & = & 
  \mu_0 \, \delta j_\varphi -
  i \, \frac{\omega}{c^2} \, \delta E_\varphi
\end{eqnarray}
The current density perturbation is
\begin{eqnarray}
  \delta j_{\rm r} & = & \rho_{\rm e} \, \delta v_{\rm r} \\
  \delta j_\varphi & = & \delta \rho_{\rm e} \, v_\varphi + 
  \rho_{\rm e} \, \delta v_\varphi
\end{eqnarray}
It is convenient to introduce a new function~$\phi$ related to the
azimuthal electric field by $\delta E_\varphi = -i \, l \, \phi / r$
(we emphasise that this function is not the scalar potential from
which the electric field could be derived from, it is just a
convenient auxiliary variable) such that the electric and magnetic
field become
\begin{eqnarray}
  \label{eq:deltaEr}
  \delta E_{\rm r} & = & - \kappa(r, \omega) \, \left[ 
    \frac{\partial\phi}{\partial r} - i \, \mu_0 \, \omega \,
    \frac{r^2}{l^2} \, \rho_{\rm e} \, \delta v_{\rm r} \right] \\
  \label{eq:deltaBz}
  \delta B_{\rm z} & = & \frac{\omega \, r}{l} \, \kappa(r, \omega) \, 
  \left[ \frac{1}{c^2} \, \frac{\partial \phi}{\partial r} - i \,
    \frac{\mu_0}{\omega} \, \rho_{\rm e} \, \delta v_{\rm r} \right]
\end{eqnarray}
We introduced the function
\begin{equation}
  \label{eq:kappa}
  \kappa(r, \omega) = \frac{1}{ 1 - ( \omega \, r / l \, c )^2}.
\end{equation}
Maxwell-Gauss equation, (\ref{eq:MaxGauss}), is therefore written
\begin{equation}
  \frac{1}{r} \, \frac{\partial}{\partial r} 
  \left( r \, \kappa(r, \omega) \, \frac{\partial \phi}{\partial r} \right) -
  \frac{l^2}{r^2} \, \phi = - \frac{\delta\rho_{\rm e}}{\varepsilon_0} \, +
  i \, \mu_0 \, \frac{\omega}{r} \, \frac{\partial}{\partial r}
  \left( \frac{r^3}{l^2} \, \kappa(r, \omega) \, \rho \, \delta v_{\rm r} \right)
\end{equation}
From the continuity equation, (\ref{eq:Continuite}), we get
\begin{equation}
  \delta \rho_{\rm e} = \frac{1}{i\, (\omega - l \, \Omega)} \, 
  \left[ \frac{1}{r} \, \frac{\partial}{\partial r} 
    ( r \, \rho_{\rm e} \, \delta v_{\rm r}) +
    i \, \frac{l}{r} \, \rho \, \delta v_\varphi \right]
\end{equation}
The electric drift approximation gives
\begin{eqnarray}
  \delta v_{\rm r} & = & \frac{\delta E_\varphi}{B_{\rm z}} 
  = - i \, \frac{l}{r} \, \frac{\phi}{B_{\rm z}} \\
  \delta v_\varphi & = & \frac{E_r \, \delta B_{\rm z} - \delta E_r \, B_{\rm z}}{B_{\rm z}^2}
\end{eqnarray}
After some algebra, the eigenvalue problem for the relativistic
diocotron instability in cylindrical geometry is written
\begin{equation}
  \label{eq:ValPropDiocRelat}
  \frac{1}{r} \, \frac{\partial}{\partial r} 
  \left[ r \, \kappa(r,\omega) \, \frac{\partial \phi}{\partial r} \right] -
  \frac{l^2}{r^2} \, [ 1 + \chi(r,\omega) ] \, \phi =
  \frac{l \, \phi}{r \, ( \omega - l \, \Omega ) } \, \kappa(r,\omega) \,
  \left[ 1 - \frac{\Omega \, r}{c} \, \frac{\omega \, r}{l \, c} \right] \,
  \frac{\partial}{\partial r} \left( \frac{\omega_{\rm p}^2}{\omega_{\rm c}} \right)
\end{equation}
with
\begin{equation}
  \chi(r,\omega) = \kappa(r, \omega) \, \frac{\omega_{\rm p}^2}{\omega_{\rm c}} \, 
  \left( \frac{r}{l \, c} \right)^2 \, 
  \left( \frac{\omega_{\rm p}^2}{\omega_{\rm c}} + 
    2 \, \kappa(r, \omega) \, \frac{\omega}{l} \right) 
\end{equation}
The eigenvalue equation~(\ref{eq:ValPropDiocRelat}) is very general.
It describes the motion of small electromagnetic perturbations around
the given equilibrium state, Eqs.~(\ref{eq:PoissonDiocRelat}) and
(\ref{eq:AmpereDiocRelat}), in the electric drift approximation
Eq.~(\ref{eq:vDiocRelat}). Many aspect of the relativistic diocotron
instability can be investigated with this eigenvalue equation.  In
order to solve the eigenvalue problem, boundary conditions need to be
imposed at the plasma/vacuum interface. They play a decisive role in
the presence or absence of the instability. How to treat these
transitions between plasma and vacuum is discussed in the next
subsection.

Note that in the non-relativistic limit, the generalised linear
eigenvalue problem, Eq.~(20) in \cite{2007A&A...464..135P}, is
recovered
\begin{equation}
  \label{eq:ValPropDiocNonRelat}
  \frac{1}{r} \, \frac{\partial}{\partial r} 
  \left[ r \, \frac{\partial \phi}{\partial r} \right] -
  \frac{l^2}{r^2} \, \phi =
  \frac{l \, \phi}{r \, ( \omega - l \, \Omega ) } \,
  \frac{\partial}{\partial r} \left( \frac{\omega_{\rm p}^2}{\omega_{\rm c}} \right)
\end{equation}
This is the standard eigenvalue problem for the non-relativistic
diocotron instability in cylindrical geometry.

\subsection{Boundary conditions}

\label{sec:Boundary}

In laboratory experiments, the plasma is usually confined between an
inner and an outer conducting wall. However, in pulsar electrospheres,
no such outer device exists to constraint the electric field at the
outer boundary. Radiation from the plasma could propagate into vacuum
to infinity, carrying energy away from the plasma by Poynting flux. To
allow for this electromagnetic wave production by the instabilities
studied in this work, the outer wall is removed. The electromagnetic
field is solved analytically in vacuum and matched to the solution in
the plasma at the plasma/vacuum interface located at $r=R_2$. First we
discuss the situation in which an outer wall exists and next consider
outgoing waves.

\subsubsection{Outer wall}

When vacuum regions exist between the plasma column and the walls,
special care is required at the sharp plasma/vacuum interfaces.
Indeed, the right-hand side of Eq.~(\ref{eq:ValPropDiocRelat}) then
involves Dirac distribution functions~$\delta(r)$ because the function
$\omega_{\rm p}^2 / \omega_{\rm c}$ is discontinuous at $R_1$ and
$R_2$. In other words, its derivative is
\begin{eqnarray}
  \label{eq:Dirac1}
  \frac{\partial}{\partial r} 
  \left( \frac{\omega_{\rm p}^2}{\omega_{\rm c}} \right) = 
  \delta ( r - R_1 ) \, \left. \frac{\partial}{\partial r} 
    \left( \frac{\omega_{\rm p}^2}{\omega_{\rm c}} \right) 
  \right|_{r=R_1} -
  \delta ( r - R_2 ) \, \left. \frac{\partial}{\partial r} 
    \left( \frac{\omega_{\rm p}^2}{\omega_{\rm c}} \right) 
  \right|_{r=R_2} +
  \left. \frac{\partial}{\partial r} 
    \left( \frac{\omega_{\rm p}^2}{\omega_{\rm c}} \right)  
    \right|_{\rm regular}
\end{eqnarray}
where $||_{\rm regular}$ means the regular (or continuous) part of the
derivative, i.e. which does not involve distribution functions. It
vanishes in the vacuum regions, $r<R_1$ and $r>R_2$. Therefore, the
first order derivative of~$\phi$ is not continuous at these
interfaces. To overcome this difficulty, we decompose the space
between the two walls into three distinct regions:
\begin{itemize}
\item region~I: vacuum space between inner wall and inner boundary of
  the plasma column, with the solution for the function~$\phi$ denoted
  by $\phi_\mathrm{I}$, defined for $W_1 \le r \le R_1$~;
\item region~II: the plasma column itself located between $R_1$ and
  $R_2$, solution denoted by $\phi_\mathrm{II}$, defined for $R_1
  \le r \le R_2$~;
\item region~III: vacuum space between the outer boundary of the
  plasma column and the outer wall, solution denoted by
  $\phi_\mathrm{III}$, defined for $R_2 \le r \le W_2$.
\end{itemize}
In regions~I and III, the vacuum solutions should satisfy the required
boundary conditions, $\phi_\mathrm{I}(W_1) = 0$ and
$\phi_\mathrm{III}(W_2) = 0$.

The jumps in the derivative $\partial\phi/\partial r$ at each
interface are easily founded from Eq.~(\ref{eq:Dirac1}). At~$R_1$, the
jump is 
\begin{equation}
  \label{eq:Jump1}
  \frac{\partial \phi_{\rm II}}{\partial r}(R_1) -
  \frac{\partial \phi_{\rm I }}{\partial r}(R_1)
  = \frac{l \, \phi(R_1)}{R_1 \, ( \omega - l \, \Omega(R_1) ) } \,
  \left( 1 - \frac{\Omega(R_1) \, R_1}{c} \, 
    \frac{\omega \, R_1}{l \, c} \right) \,
  \frac{\omega_{\rm p}^2(R_1)}{\omega_{\rm c}(R_1)} .
\end{equation}
Similarly, at the outer interface at~$R_2$, we obtain,
\begin{equation}
  \label{eq:Jump2}
  \frac{\partial \phi_{\rm III}}{\partial r}(R_2) -
  \frac{\partial \phi_{\rm II }}{\partial r}(R_2)
  = - \frac{l \, \phi(R_2)}{R_2 \, ( \omega - l \, \Omega(R_2) ) } \,
  \left( 1 - \frac{\Omega(R_2) \, R_2}{c} \, 
    \frac{\omega \, R_2}{l \, c} \right) \,
  \frac{\omega_{\rm p}^2(R_2)}{\omega_{\rm c}(R_2)} .
\end{equation}

\subsubsection{Outgoing wave solution}

Because of the wall located at~$r=W_2$, the outer boundary condition
$\phi_\mathrm{III}(W_2) = 0$ enforces $E_\varphi(W_2)=0$.  It
therefore prevents escaping waves from the system due to the vanishing
outgoing Poynting flux, $E_\varphi\,B_z/\mu_0=0$. In pulsar
magnetospheres, no such wall exists. So, in order to let the system
produce outgoing electromagnetic waves, we remove the outer wall in
this case and solve the vacuum wave equation for~$\phi$ which then
reads
\begin{equation}
  \label{eq:OndePhiVide}
  \frac{1}{r} \, \frac{\partial}{\partial r} 
  \left[ r \, \kappa(r,\omega) \, \frac{\partial \phi}{\partial r} \right] -
  \frac{l^2}{r^2} \, \phi = 0
\end{equation}
This equation can also be derived directly from the vector wave
equation
\begin{equation}
  \label{eq:VectorWave}
  \Delta \vec{E} - \frac{1}{c^2} \, 
  \frac{\partial^2 \vec{E}}{\partial t^2} = \vec{0}
\end{equation}
projected along the $e_\varphi$ axis.  To find the right outgoing wave
boundary conditions, it is therefore necessary to solve the vector
wave equation in cylindrical coordinates using vector cylindrical
harmonics as described for instance in \cite{Stratton1941} and
\cite{1953mtp..book.....M}. The solutions for the function~$\phi$ to
be an outgoing wave in vacuum outside the plasma column and which
vanishes at infinity is given by (region~III with $W_2 = +\infty$)
\begin{equation}
  \label{eq:PhiOutgoing}
  \phi_{\rm III} = K \, r \,\frac{\partial}{\partial r} 
  H_l \left( \frac{\omega\,r}{c} \right) =
  K \, \frac{\omega \, r}{c} \, H_l' \left( \frac{\omega\,r}{c} \right)
\end{equation}
where the cylindrical outgoing wave function is given by $H_l$
\citep{Stratton1941}, the Hankel function of first kind and of
order~$l$ related to the Bessel functions by $H_l(x) = J_l(x) + i \,
Y_l(x)$, \citep{1965hmfw.book.....A}. The prime~$'$ means derivative
of the function evaluated at the point given in parentheses. $K$ is a
constant to be determined from the boundary condition at $R_2$.
Eliminating the constant~$K$, we conclude that the boundary condition
to impose on $\phi$ is
\begin{equation}
  \label{eq:CLPhiR2}
  \left[ H_l'\left( \frac{\omega\,R_2}{c} \right) + 
    \frac{\omega\,R_2}{c} \, H_l''\left( \frac{\omega\,R_2}{c} \right) \right] 
  \, \phi_{\rm III}(R_2) -
  R_2 \, H_l'\left( \frac{\omega\,R_2}{c} \right) \, 
  \frac{\partial\phi_{\rm III}}{\partial r}(R_2) = 0
\end{equation}
The boundary conditions {\it expressed in region~II for~$\phi_{\rm
    II}$} are found by replacing $\phi_{\rm III}'(R_2)$ from
Eq.~(\ref{eq:Jump2}), and recalling that~$\phi$ is continuous,
therefore $\phi_{\rm III}(R_2) = \phi_{\rm II}(R_2) =
\phi_{\rm}(R_2)$. We find
\begin{eqnarray}
  \label{eq:Jump3}
  & & \left[ H_l'\left( \frac{\omega\,R_2}{c} \right) + 
    \frac{\omega\,R_2}{c} \, H_l''\left( \frac{\omega\,R_2}{c} \right) \right] 
  \, \phi(R_2) - \\
  & & R_2 \, H_l'\left( \frac{\omega\,R_2}{c} \right) \,
  \left[ \frac{\partial\phi_{\rm II }}{\partial r}(R_2) - 
    \frac{l \, \phi(R_2)}{R_2 \, ( \omega - l \, \Omega(R_2) ) } \,
    \left( 1 - \frac{\Omega(R_2) \, R_2}{c} \, 
      \frac{\omega \, R_2}{l \, c} \right) \,
    \frac{\omega_{\rm p}^2(R_2)}{\omega_{\rm c}(R_2)} \right] = 0 \nonumber
\end{eqnarray}

\subsection{Algorithm}
\label{sec:Algo}

The eigenvalue problem, Eq.~(\ref{eq:ValPropDiocRelat}), is solved by
standard numerical technics. We have implemented a shooting method as
follows.

First, we guess a good initial value for the eigenvalue~$\omega$.
Then, the ordinary differential equation,
Eq.~(\ref{eq:ValPropDiocRelat}), is integrated numerically from $W_1$
to $W_2$ with a fifth-order Runge-Kutta or a Bulirsch-Stoer method.
More precisely, at $r=W_1$, the initial conditions are, $\phi=0$ and
$\partial\phi/\partial r = 1$.  Integration is performed in region~I
until the first vacuum/plasma interface is reached at $r=R_1$. There,
the first order derivative in $\phi$ is subject to a discontinuity
given by the jump in Eq.~(\ref{eq:Jump1}). Knowing $\partial\phi_{\rm
  I}/\partial r(R_1)$, we deduce $\partial\phi_{\rm II}/\partial
r(R_1)$. Therefore, the integration is continued in region~II with the
initial conditions, $\phi_{\rm II}(R_1) = \phi_{\rm I}(R_1)$ (because
$\phi$ is continuous) and $\partial\phi_{\rm II}/\partial r(R_2)$
until the second vacuum/plasma interface is reached at $R_2$. The
first order derivative in $\phi$ is now subject to another
discontinuity given by the jump in Eq.~(\ref{eq:Jump2}).  Knowing
$\partial\phi_{\rm II}/\partial r(R_2)$, we deduce $\partial\phi_{\rm
  III}/\partial r(R_2)$.  Integration is continued in region~III with
the initial conditions, $\phi_{\rm III}(R_2) = \phi_{\rm II}(R_2)$ and
$\partial\phi_{\rm III}/\partial r(R_2)$ until the endpoint $W_2$. At
the end of the process, the function~$\phi_{\rm III}$ does not
necessarily satisfy the desired boundary conditions. Indeed, the
eigenvalue is found whenever the function~$\phi$ at the outer wall
vanishes $\phi_{\rm III}(W_2) = 0$.  Finding $\omega$ is therefore
equivalent to finding the root of $\phi_{\rm III}(W_2)$ with respect
to the eigenvalue~$\omega$.

For the pulsar electrosphere, the situation is very similar except
that no calculation is performed in region~III. The boundary condition
for outgoing waves is applied at $R_2$, see Eq.~(\ref{eq:Jump3}).
Actually for pulsars, we compare both boundary conditions.

\section{Algorithm check}
\label{sec:Check}

In order to check our algorithm in different configurations, we
compute the eigenvalues for both a non-relativistic cylindrical plasma
column and a relativistic planar diode geometry. For some special
density profiles, the exact analytical dispersion relations are known
and used for comparison with the numerical results.

\subsection{Non-relativistic plasma column}
\label{sec:ColNRel}

In cylindrical geometry, an exact analytical solution for the
dispersion relation can be found in the non-relativistic case,
\citep{Davidson1990}.  We use these results to check our algorithm in
cylindrical coordinates.

The magnetic field is constant and uniform in the whole space, $B_{\rm
  z}(r) = B_0$. We do not need to solve Maxwell-Amp\`ere equation
because the magnetic perturbation is neglected in the non-relativistic
limit.  The particle number density and charge density are also
uniform in the whole plasma column such that
\begin{equation}
  \label{eq:RhoDiocRect}
  \rho_{\rm e}(r) = \left \lbrace
    \begin{array}{lcl}
      0   & , & W_1 \le r \le R_1 \\
      \rho_0 = {\rm const} & , & R_1 \le r \le R_2 \\
      0   & , & R_2 \le r \le W_2
    \end{array}
  \right.
\end{equation}
Solving Maxwell-Gauss equation~(\ref{eq:MaxGauss}), the equilibrium
radial electric field is
\begin{equation}
  \label{eq:ErDiocRect}
  E_{\rm r}(r) = \left \lbrace
    \begin{array}{lcl}
      0   & , & W_1 \le r \le R_1 \\
      \displaystyle{ \frac{\rho_0 \, r}{2 \, \varepsilon_0} \, 
        \left( 1 - \frac{R_1^2}{r^2} \right) }
      & , & R_1 \le r \le R_2 \\
      \displaystyle{ \frac{\rho_0}{2 \, \varepsilon_0} \, 
        \frac{R_2^2 - R_1^2}{r} }
      & , & R_2 \le r \le W_2
    \end{array}
  \right.
\end{equation}
and the corresponding electric drift speed in the plasma
\begin{equation}
  \label{eq:DiocotronFreq}
  \Omega = - \omega_{\rm D} \, \left( 1 - \frac{R_1^2}{r^2} \right) .
\end{equation}
The diocotron frequency is constant, $\omega_{\rm D} = \rho_0 / 2 \,
\varepsilon_0 \, B_0 = {\rm const}$.  The solutions to the dispersion
relation for this particular case are
\begin{equation}
  \label{eq:ValPropNRel}
  \omega = - \frac{\omega_{\rm D}}{2} \,
  \left[ b_l \pm \sqrt{b_l^2 - 4 \, c_l} \right]
\end{equation}
The coefficients $b_l$ and $c_l$ are given by
\begin{eqnarray}
  \label{eq:b_l}
  b_l & = & \left\{ l \, \left[ 1 - \left( \frac{R_1}{R_2} \right)^2 \right] \, 
    \left[ 1 - \left( \frac{W_1}{W_2} \right)^{2l} \right] +
    \left[ 1 - \left( \frac{R_1}{R_2} \right)^{2l} \right] \, 
    \left[ \left( \frac{R_2}{W_2} \right)^{2l} - 
      \left( \frac{W_1}{R_1} \right)^{2l} \right] \right\} \,
  \left[ 1 - \left( \frac{W_1}{W_2} \right)^{2l} \right]^{-1} \\
  \label{eq:c_l}
  c_l & = & \left\{ l \, \left[ 1 - \left( \frac{R_1}{R_2} \right)^2 \right] \,
    \left[ 1 - \left( \frac{R_1}{W_2} \right)^{2l} \right] \,
    \left[ 1 - \left( \frac{W_1}{R_1} \right)^{2l} \right] - \right. \nonumber \\
  & & \left. \left[ 1 - \left( \frac{R_2}{W_2} \right)^{2l} \right] \, 
    \left[ 1 - \left( \frac{W_1}{R_1} \right)^{2l} \right] \,
    \left[ 1 - \left( \frac{R_1}{R_2} \right)^{2l} \right] \right\} \,
  \left[ 1 - \left( \frac{W_1}{W_2} \right)^{2l} \right]^{-1}
\end{eqnarray}
A sample of eigenvalues is shown in Tab.~\ref{tab:DiocCylNRel}, for
$W_1 = 1$ and $W_2 = 10$ and different aspect ratios, $d_1 = R_1 /
W_2$ and $d_2 = R_2 / W_2$. The exact analytical solution,
Eq.~(\ref{eq:ValPropNRel}), is compared with the numerical solution.
The relative errors in the real and imaginary part of the eigenvalues
are
\begin{eqnarray}
  \label{eq:Erreur}
  \varepsilon_{\rm Re} & = & 
  \left| \frac{{\rm Re} \, (\omega) - {\rm Re} \, (\omega_{\rm exact})}
    {{\rm Re} \, (\omega_{\rm exact})} \right| \le 3.2 \times 10^{-11} \\
  \varepsilon_{\rm Im} & = & 
  \left| \frac{{\rm Im} \, (\omega) - {\rm Im} \, (\omega_{\rm exact})}
    {{\rm Im} \, (\omega_{\rm exact})} \right| \le 1.1 \times 10^{-10}
\end{eqnarray}
The precision is excellent, it reaches 10~digits. Our algorithm
computes quickly and accurately the eigenvalues in cylindrical
geometry with vacuum gaps between the plasma column and the walls.
The eigenvalues obtained in this example are good initial guesses to
study the relativistic problem in the low speed limit.  Next, we turn
to the relativistic planar case.
\begin{table}[htbp]
  \centering
  \begin{tabular}{cccccc}
    \hline
    mode $l$ & $d_1$ & $d_2$ & $\omega_{\rm num}$ & 
    $\varepsilon_{\rm Re}$ & $\varepsilon_{\rm Im}$ \\
    \hline
    2 & 0.4 & 0.5 & -3.772e-01 + 7.176e-02 \, i & 2.046e-11 & 3.148e-11 \\
    3 & 0.4 & 0.5 & -5.456e-01 + 2.267e-01 \, i & 3.330e-15 & 7.352e-15 \\
    4 & 0.4 & 0.5 & -7.216e-01 + 2.988e-01 \, i & 7.845e-12 & 1.855e-12 \\
    5 & 0.7 & 0.9 & -1.147e+00 + 5.787e-02 \, i & 6.050e-12 & 2.618e-11 \\
    7 & 0.6 & 0.7 & -9.315e-01 + 3.307e-01 \, i & 1.006e-11 & 3.832e-12 \\
    \hline
  \end{tabular}
  \caption{Numerical eigenvalues~$\omega_{\rm num}$ and relative errors
    for the non-relativistic plasma column for different mode~$l$ and 
    different aspect ratios, $d_1 = R_1 / W_2$, and $d_2 = R_2 / W_2$ 
    with $W_1 = 1$ and $W_2 = 10$. 
    The precision of the computed eigenvalues reaches 10~digits.}
  \label{tab:DiocCylNRel}  
\end{table}

\subsection{Relativistic planar diode geometry}

\begin{figure}[htbp]
  \centering
  \begin{tabular}{cc}
  \includegraphics[scale=0.6]{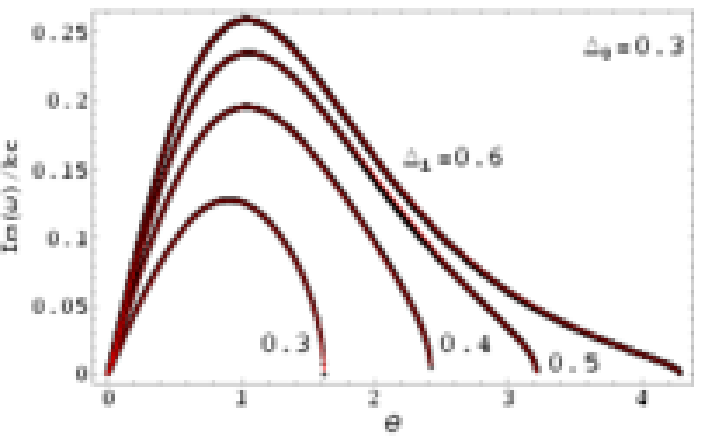} &
  \includegraphics[scale=0.6]{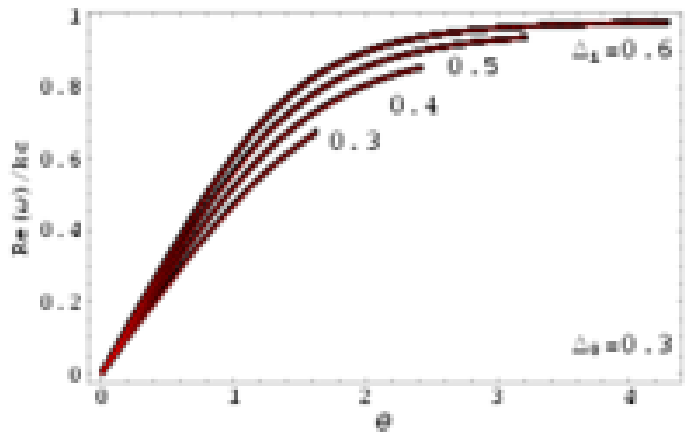}
  \end{tabular}
  \caption{Imaginary part, on the left panel, and real part, 
    on the right panel, of the eigenfrequencies, (in black dots),
    compared to the analytical exact dispersion relation in the long
    wavelength limit Eq.~(\ref{eq:DispRelatCart}) and (\ref{eq:ImOm}),
    (red curves). The parameters are $\Delta_0=0.3$ and
    $\Delta_i=0.3,0.4,0.5,0.6$.}
  \label{fig:David1}
\end{figure}

\begin{figure}[htbp]
  \centering
  \begin{tabular}{cc}
  \includegraphics[scale=0.6]{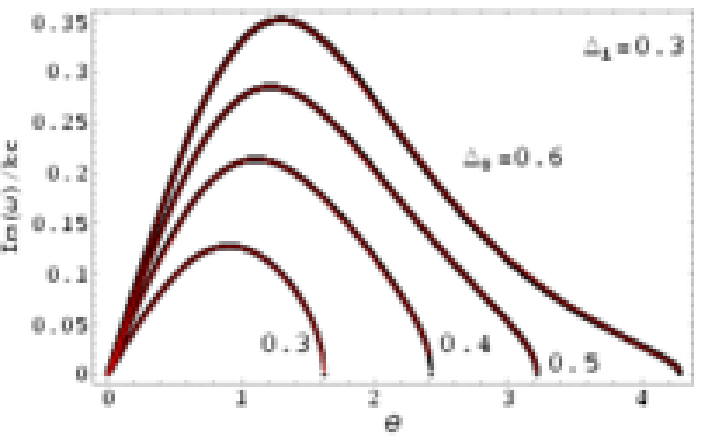} &
  \includegraphics[scale=0.6]{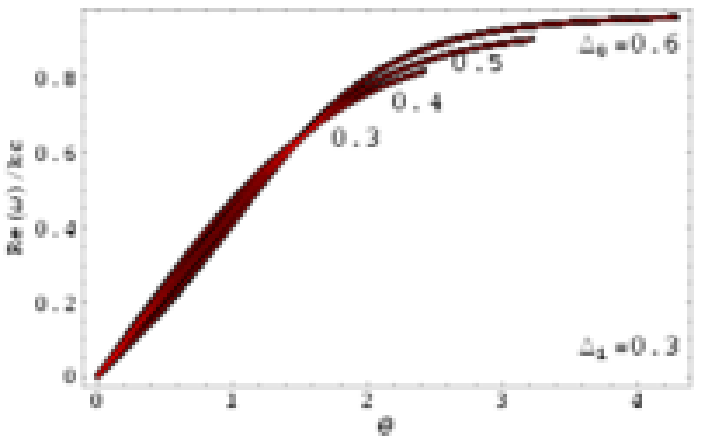}
  \end{tabular}
  \caption{Imaginary part, on the left panel, and real part, 
    on the right panel, of the eigenfrequencies, (in black dots),
    compared to the analytical exact dispersion relation in the long
    wavelength limit Eq.~(\ref{eq:DispRelatCart}) and (\ref{eq:ImOm}),
    (red curves). The parameters are $\Delta_i=0.3$ and
    $\Delta_0=0.3,0.4,0.5,0.6$.}
  \label{fig:David2}
\end{figure}

\begin{figure}[htbp]
  \centering
  \begin{tabular}{cc}
  \includegraphics[scale=0.6]{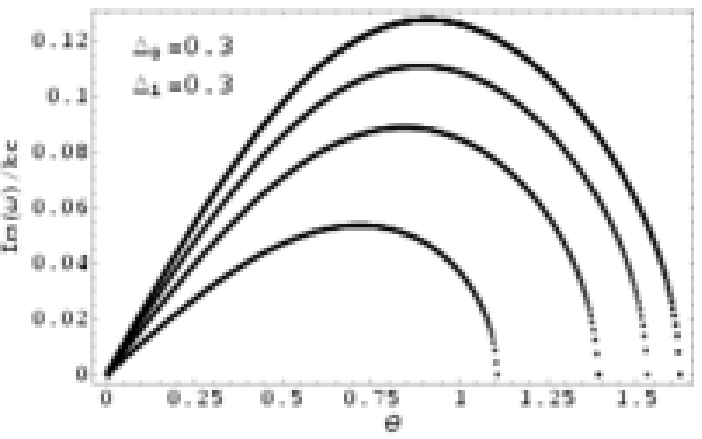} &
  \includegraphics[scale=0.6]{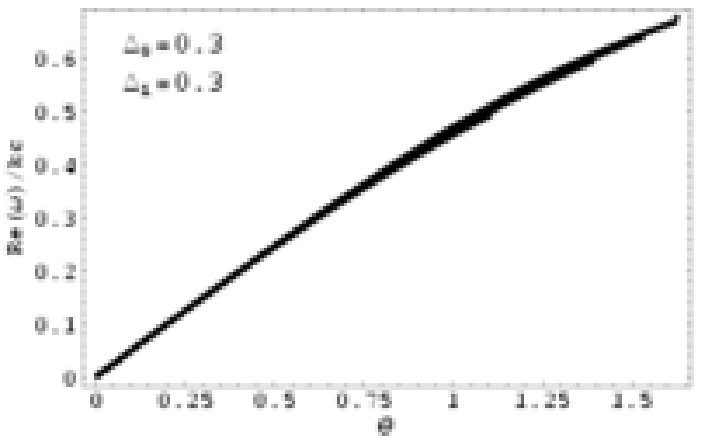}
  \end{tabular}
  \caption{Imaginary part, on the left panel, and real part,
    on the right panel, of the eigenfrequencies for arbitrary
    wavelength, $k\,d$. The parameters are $\Delta_i=0.3$ and
    $\Delta_0=0.3$. Check with Fig.6 of \cite{1988PhFl...31.1727D}.}
  \label{fig:David3}
\end{figure}

\begin{figure}[htbp]
  \centering
  \begin{tabular}{cc}
  \includegraphics[scale=0.6]{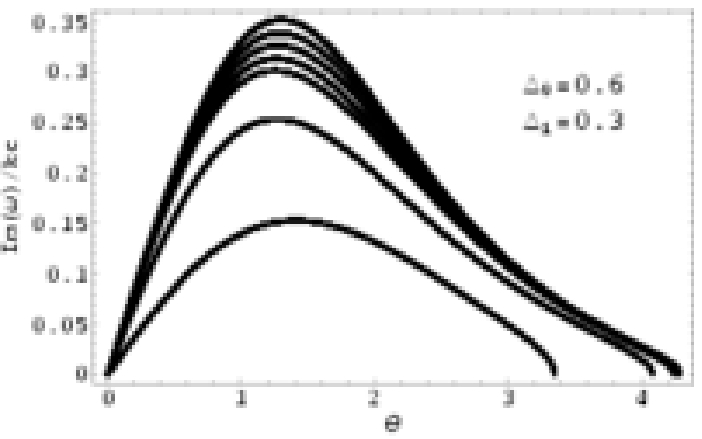} &
  \includegraphics[scale=0.6]{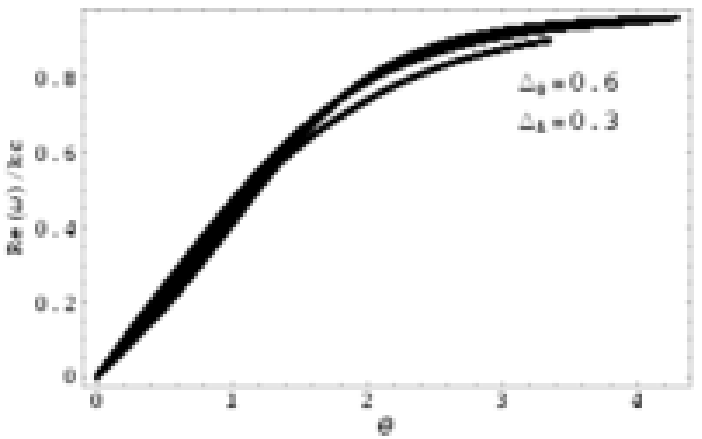}
  \end{tabular}
  \caption{Imaginary part, on the left panel, and real part,
    on the right panel, of the eigenfrequencies for arbitrary
    wavelength, $k\,d$. The parameters are $\Delta_i=0.3$ and
    $\Delta_0=0.6$. Check with Fig.7 of \cite{1988PhFl...31.1727D}.}
  \label{fig:David4}
\end{figure}

In the relativistic planar diode geometry, the dispersion relation is
also known analytically in the long wavelength limit. Thus it is an
interesting case to check our algorithm for the relativistic diocotron
instability. For completeness, we recall the main results. For a
detailed discussion, see \cite{1987PhLA..125...61D,
  1988PhFl...31.1727D}. The plasma is drifting in the $y$-direction at
a speed~$V_{\rm y}$ and located between $x=X_1$ and $x=X_2$. The
cathode is located at $x=0$ and the anode at $x=d$. We use Cartesian
coordinates $(x,y,z)$.  The eigenvalue problem in Cartesian
coordinates for the diocotron instability in a relativistic planar
diode geometry is
\begin{equation}
  \label{eq:EigValuePlanar}
  \frac{\partial^2 \phi}{\partial x^2} - k^2 \, 
  \left( 1 - \frac{\omega^2}{k^2 \, c^2} +
    \frac{\omega_{\rm Dc}^2}{k^2 \, c^2} \right) \, \phi
  = - \frac{k \, \phi}{\omega - k \, V_{\rm y}} \, 
  \left( 1 - \frac{V_{\rm y}}{c} \, \frac{\omega}{k\,c} \right) \,
  \frac{\partial \omega_{\rm Dc}}{\partial x}
\end{equation}
where $k$ is the wavenumber.  In this paragraph, because of the
Cartesian geometry, the diocotron frequency is defined as (no factor
$1/2$)
\begin{equation}
  \label{eq:DiocFreqCart}
  \omega_{\rm Dc} = \frac{\omega_{\rm p}^2}{\omega_{\rm c}}
\end{equation}
The applied external magnetic field is constant and uniform outside
the plasma layer, $B_{\rm z}(r) = B_0$ for $x \ge X_2$.  The particle
number density is chosen such that
\begin{equation}
  \label{eq:DiocRelRect}
  \frac{n(x)}{\gamma(x)} = \left \lbrace
    \begin{array}{lcl}
      0   & , & W_1 \le x \le X_1 \\
      \displaystyle{\frac{n_0}{\gamma_0}} = {\rm const} & , & X_1 \le x \le X_2 \\
      0   & , & X_2 \le x \le W_2
    \end{array}
  \right.
\end{equation}
The Lorentz factor is $\gamma = 1 / \sqrt{1 - V_{\rm y}^2 / c^2}$.
For this particular choice of density profile, the diocotron frequency
is constant throughout the layer cross section, $\omega_{\rm Dc} = {\rm
  const}$.  Moreover, the drift speed is
\begin{eqnarray}
  \label{eq:DriftSpeedCart}
  V_{\rm y} & = & c \, \tanh \, [ \omega_{\rm Dc} \, ( x - X_1 ) / c ]
\end{eqnarray}
In the long-wavelength perturbation limits, corresponding to $k\,d \ll
1$, the dispersion relation reads
\begin{eqnarray}
  \label{eq:DispRelatCart}
  \frac{{\rm Re}(\omega)}{k\,c} & = & \frac{1}{2} \,
  \left[ \sinh \theta + 2 \, \frac{\Delta_i}{\Delta_b} \, \theta \, \cosh \theta \right]
  \left[ \cosh \theta + \frac{\theta}{\Delta_b \, \sinh \theta} \, 
    ( \Delta_i \, \cosh^2 \theta + \Delta_0 ) \right]^{-1} \\
  \label{eq:ImOm}
  \frac{{\rm Im}(\omega)}{k\,c} & = & \frac{\theta}{2} \,
  \sqrt{ p^2 - \frac{\sinh^2 \theta}{\theta^2} }
  \left[ \cosh \theta + \frac{\theta}{\Delta_b \, \sinh \theta} \, 
    ( \Delta_i \, \cosh^2 \theta + \Delta_0 ) \right]^{-1}
\end{eqnarray}
where the quantities are defined by
\begin{eqnarray}
  \Delta_i & = &         X_1   / W_2 \\
  \Delta_b & = & ( X_2 - X_1 ) / W_2 \\
  \Delta_0 & = & ( W_2 - X_2 ) / W_2 \\
  p & = & 2 \, \sqrt{\Delta_i \, \Delta_0} / \Delta_b \\
  \theta & = & \omega_{\rm Dc} \, ( X_2 - X_1 ) / c .
\end{eqnarray}
Results for the eigenvalues are shown in Figs.~\ref{fig:David1} and
\ref{fig:David2} in the long wavelength limit and compared with the
exact dispersion relations, Eqs.~(\ref{eq:DispRelatCart}) and
(\ref{eq:ImOm}). The relativistic diocotron instability for arbitrary
wavelength is shown in Figs.~\ref{fig:David3} and \ref{fig:David4}.
Our numerical results agree with very good accuracy to those found by
\cite{1987PhLA..125...61D, 1988PhFl...31.1727D} who directly solved
for the dispersion relation in the general case, i.e. for arbitrary
$k\,d$.


\section{RESULTS}
\label{sec:Results}

We demonstrated that our numerical algorithm gives accurate results in
the non-relativistic cylindrical geometry as well as in the
relativistic regime in Cartesian coordinates. In this section, we
compute the eigenspectra of the relativistic diocotron instability in
cylindrical coordinates, for various equilibrium density, electric
field, and velocity profiles.  Application to pulsar's electrosphere
is also discussed.

\subsection{Plasma column}
\label{subsec:Lab}

First, we consider the laboratory plasma in cylindrical geometry,
confined by some external experimental electromagnetic device. The
external applied magnetic field and the density profile are specified
as initial data. We use the simple profile presented in
Sect.~\ref{sec:ProfDens}.

In the non-relativistic limit, a very good first guess is given by
Eq.~(\ref{eq:ValPropNRel}). The influence of the relativistic and
electromagnetic effects are then investigated by slowly increasing the
maximal speed at the outer edge of the plasma column, $v_{\rm max} =
R_2 \, \Omega(R_2)$.  

In the non-relativistic flow, the growth rate of the diocotron
instability only depends on the ratios~$d_1 = R_1/W_2$, $d_2 =
R_2/W_2$ and $w = W_1/W_2$. In order to avoid variation of the growth
rate or stabilisation of the instability purely due to geometrical
effects, these ratios are kept constant while $v_{\rm max}$ is
increased. To this end, for instance, we increase $W_2$ and adjust the
other parameters $R_1,R_2,W_1$ to fit the constant ratios conditions,
$w=cst, d_1=cst, d_2=cst$.

\begin{figure}[htbp]
  \centering
  \begin{tabular}{cc}
  \includegraphics[scale=0.6]{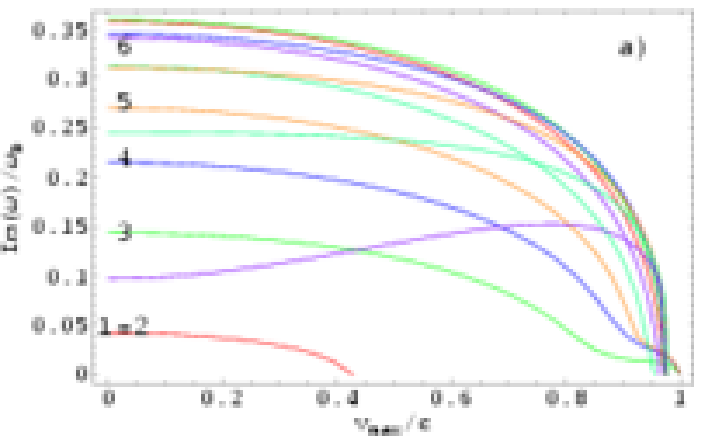} &
  \includegraphics[scale=0.6]{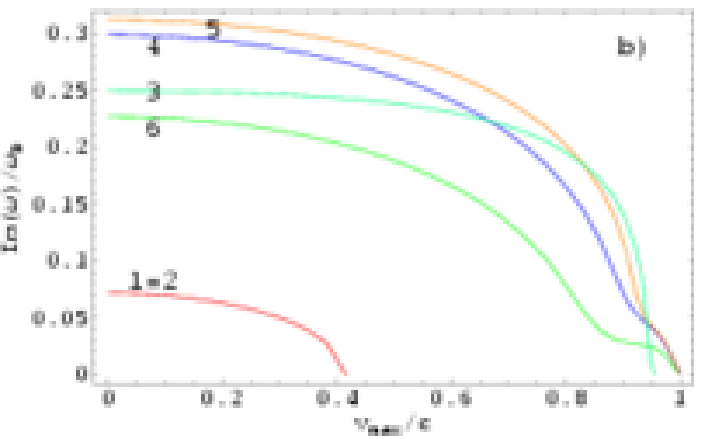}
  \end{tabular}
  \caption{Stabilisation of the diocotron instability when
    relativistic and electromagnetic effects are included.  The
    relativistic diocotron frequency is constant in the whole plasma
    column, $\omega_{\rm D} = {\rm cste}$. The geometric aspect ratios
    are, $w=0.1$, $d_1=0.45$ and $d_2=0.5$ for Fig.~a) and $w=0.1$,
    $d_1=0.4$ and $d_2=0.5$ for Fig.~b). Only the first unstable modes
    are labeled, $l=2,3,4,5,6$ to avoid overloading the plots. Each
    coloured curve depicts a different azimuthal mode number~$l$.}
  \label{fig:ColRel1}
\end{figure}

Two cases are presented in Fig.~\ref{fig:ColRel1}. The first one has
$w=0.1$, $d_1=0.45$ and $d_2=0.5$, Fig.~\ref{fig:ColRel1}a), whereas
the second one has $w=0.1$, $d_1=0.4$ and $d_2=0.5$,
Fig.~\ref{fig:ColRel1}b). The eigenvalues are conveniently normalised
to the diocotron frequency~$\omega_\mathrm{D}$. For non-relativistic
speeds, $v_{\rm max} \ll c$, the eigenvalues of
Sect.~\ref{sec:ColNRel} are recovered, see
Table~\ref{tab:FreqDiocCste}. The thinner the plasma layer, the larger
the number of unstable modes, respectively 5 and 12 unstable modes. In
both cases, the growth rate starts to be altered whenever $v_{\rm
  max}/c \gtrsim 0.1$.  We only labeled the first 5 unstable modes
$l=2,3,4,5,6$ in order to avoid overloading the figure.

\begin{table}[htbp]
  \centering
  \caption{Eigenvalues of the plasma column for the density profile 
    in Sect.~\ref{sec:ColNRel}. Comparison of the non-relativistic 
    one~$\omega_{\rm nrel}$ and the relativistic
    one~$\omega_{\rm rel}$ in the 
    low speed limit, $v_{\rm max} \ll c$ for $w=0.1$, $d_1=0.4$ and $d_2=0.5$.}
  \begin{tabular}{ccc}
    \hline
    mode~$l$ & $\omega_{\rm nrel}$ & $\omega_{\rm rel}$ \\
    \hline
    \hline
    2 & 3.772986e-01 + 7.176435e-02 \, i & 3.773015e-01 + 7.174847e-02 \, i \\ 
    3 & 5.456744e-01 + 2.267241e-01 \, i & 5.456773e-01 + 2.267196e-01 \, i \\
    4 & 7.216191e-01 + 2.988911e-01 \, i & 7.216219e-01 + 2.988890e-01 \, i \\
    5 & 9.004354e-01 + 3.118851e-01 \, i & 9.004382e-01 + 3.118852e-01 \, i \\
    6 & 1.080114e+00 + 2.495296e-01 \, i & 1.080116e+00 + 2.495328e-01 \, i \\
    \hline
  \end{tabular}
  \label{tab:FreqDiocCste}
\end{table}
In any case, for very high speeds, $v_{\rm max}/c \approx 1$, all the
diocotron modes become stable because the growth rate vanishes,
Fig.~\ref{fig:ColRel1}. The stabilisation process already observed in
the relativistic planar diode is not altered by the cylindrical
geometry.

\subsection{Relativistic planar diode limit}

The influence of the curvature is also studied by taking the limit of
the planar diode geometry. The curvature of the plasma column is then
increased to investigate the evolution of the growth rates.

The effective aspect ratio of the plasma layer is conveniently
described by the parameter
\begin{equation}
  \label{eq:AspectA}
  A = \frac{R_1}{R_2 - R_1}
\end{equation}
Using the equilibrium electric field profile indicated in
Sect.~\ref{sec:Electric}, in the limit of small curvature corresponding
to large aspect ratio, $A\rightarrow+\infty$, the eigenvalue problem and
equilibrium configuration is described by the relativistic planar
diode.

\begin{figure}[htbp]
  \centering
  \begin{tabular}{cc}
    \includegraphics[scale=0.6]{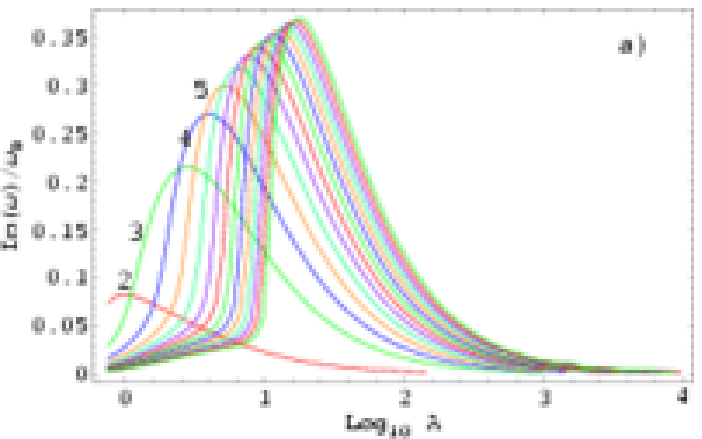} &
    \includegraphics[scale=0.6]{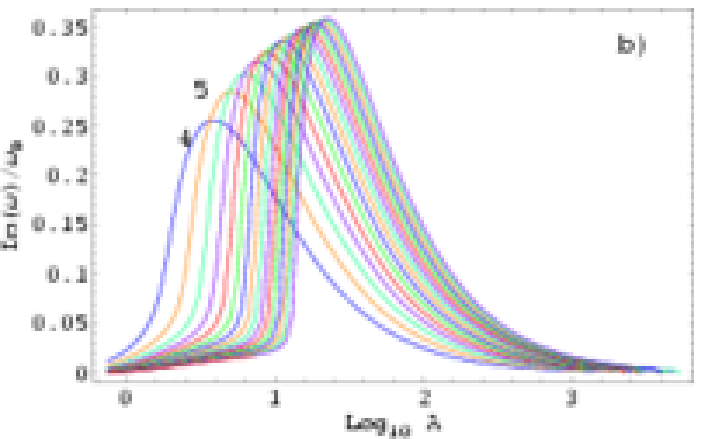}
  \end{tabular}
  \caption{Effect of the cylindrical geometry on the growth rate of the
    relativistic diocotron instability. The growth rate are normalised
    to the diocotron frequency and plotted versus the aspect
    ratio~$A$. Non-relativistic regime, $v_{\rm max} = 10^{-3}$ in
    Fig.~a) and relativistic speed, $v_{\rm max} = 0.3$ in Fig.~b).
    Only the first unstable modes are labeled, $l=2,3,4,5$ to avoid
    overloading the plots. Each coloured curve depicts a different
    azimuthal mode number~$l$.}
  \label{fig:Courbure}
\end{figure}

We show the evolution of the growth rate in the non relativistic
limit, $v_{\rm max} = 10^{-3}$, Fig.~\ref{fig:Courbure}a), and in the
relativistic case, $v_{\rm max} = 0.3$, Fig.~\ref{fig:Courbure}b).
The eigenvalues are conveniently normalised to the value of the
diocotron frequency at the outer boundary of the plasma column,
$\omega_\mathrm{D}(R_2)$. The aspect ratio has a drastic influence on
the growth rate. For large values of~$A\gg1$, all unstable modes are
stabilised, in both non-relativistic and relativistic flows. Note
however, that in the relativistic regime, the mode $l=2$ and $l=3$
have already been stabilised, whatever the aspect ratio.

When the aspect ratio~$A$ is increased, the fastest growing unstable
mode is shifting to higher azimuthal numbers~$l$ whereas the lower
azimuthal numbers~$l$ start to stabilise. Indeed, to compute the
eigenvalues for different aspect ratios, we increase $R_1$ and $R_2$
while keeping the difference $\Delta=R_2-R_1$ constant. Assuming that
the geometrical size of the perturbation fits into the layer
thickness~$\Delta$, we get an estimate of the allowed azimuthal
numbers such that
\begin{equation}
  \label{eq:Fitl}
  \frac{2\,\pi}{l} \, R_1 \approx \Delta 
\end{equation}
Therefore the fastest mode numbers are roughly $l \propto A$ and
linearly growing with the aspect ratio.

\subsection{Electrosphere}
\label{subsec:Pulsar}

\begin{figure}[htbp]
  \centering
  \begin{tabular}{cc}
  \includegraphics[scale=0.7]{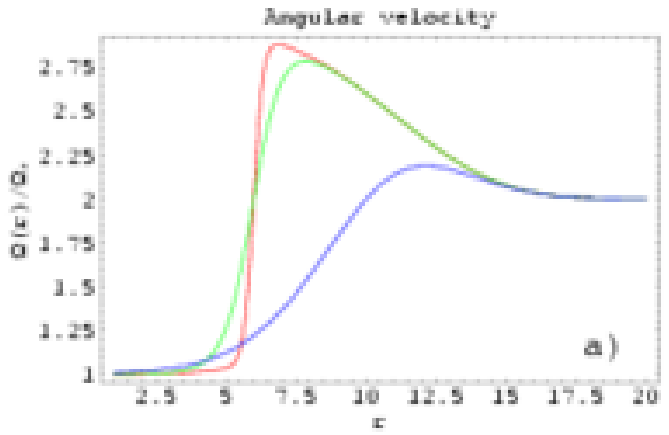} &
  \includegraphics[scale=0.7]{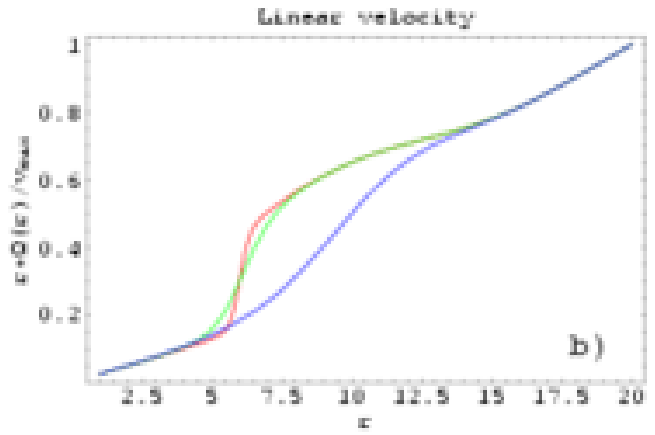}
  \end{tabular}
  \caption{Three choices of differential rotation curves
    in the plasma column for the cylindrical pulsar electrosphere,
    $\Omega_1$ in red, $\Omega_2$ in green and $\Omega_3$ in blue,
    Fig.~a). The linear speed is depicted in Fig.~b) in units of
    $v_{\rm max} = R_2 \, \Omega(R_2)$.}
  \label{fig:OmegaElec}
\end{figure}

The electrospheric non-neutral plasma, as already proved in previous
works by \cite{1985MNRAS.213P..43K} and \cite{2002A&A...384..414P}, is
confined by the rotating magnetised neutron star.  The most important
feature is the velocity profile in the plasma column.  For simplicity,
here, we assume that no vacuum gaps exist between the plasma and the
walls, so that $W_1=R_1$ and $W_2=R_2$. The rotation profile is chosen
to mimic the rotation curve obtained in the 3D electrosphere.  To
study the influence of the relativistic effects, we take the same
profiles as those given in \cite{2007A&A...464..135P}. We remind that
three different analytical expressions for the radial dependence of
$\Omega$ are chosen by mainly varying the gradient in differential
shear as follows
\begin{equation}
  \label{eq:ProfilVit}
  \Omega(r) = \Omega_* \, (2 + \tanh[  \alpha \, ( r - r_0 ) ] \, e^{-\beta\,r^4})
\end{equation}
The values used are listed in Table~(\ref{tab:Vitesse}).
\begin{table}[htbp]
  \centering
  \begin{tabular}{cccc}
    \hline
    $\Omega$ & $\alpha$ & $\beta$ & $r_0$ \\
    \hline
    $\Omega_1$ & 3.0 & $5\times10^{-5}$ & 6.0 \\
    $\Omega_2$ & 1.0 & $5\times10^{-5}$ & 6.0 \\
    $\Omega_3$ & 0.3 & $5\times10^{-5}$ & 10.0 \\
    \hline
  \end{tabular}
  \caption{Parameters for the three rotation profiles used
    to mimic the azimuthal velocity of the plasma 
    in the electrospheric disk.}
  \label{tab:Vitesse}
\end{table}
The angular velocity starts from corotation with the star $\Omega =
\Omega_*$, followed by a sharp increase around $r=6$ for
$\Omega_{1,2}$, and a less pronounced gradient around $r=10$ for
$\Omega_3$.  Finally the rotation rate asymptotes twice the neutron
star rotation speed, Fig.~\ref{fig:OmegaElec}.
The results of \cite{1985MNRAS.213P..43K} and
\cite{2002A&A...384..414P} have been computed for non-relativistic
speeds. However, confinement of non-neutral plasmas by some external
electromagnetic fields is very general and also applies when
relativistic effects are taken into account,
\cite{1986PhRvA..33.4284T}.  In case of relativistic motions within
the electrosphere, we would expect some quantitative changes in the
rotation curves because of the relativistic factor appearing in the
expression for the charge density (due to the current displacement)
\begin{equation}
  \label{eq:Chargedensity}
  \rho = - 2 \, \varepsilon_0 \, \frac{\vec{\Omega} \cdot \vec{B}}
  { 1 - r^2 \, \Omega^2 / c^2 }
\end{equation}
Nevertheless, differential rotation is still expected, even in this
latter case. That is why we took typical rotation curves as those
depicted in Fig.~\ref{fig:OmegaElec}~a). We emphasise that the
electromagnetic field as well as the charge density resulting from the
chosen rotation profiles are determined in a {\it full
  self-consistent} manner, as described in Sect.~\ref{sec:Setup}.
Relativistic speeds are only reached in the extended part of the
electrosphere, i.e. the outer part approaching the light cylinder.  In
Fig.~\ref{fig:OmegaElec}~b), the linear speed in the disk is
plotted~$r\,\Omega(r)/v_{\rm max}$, the highest speed~$v_{\rm max}<c$
corresponds to the largest radii~$r$, i.e. the outer part of the disk.
Nevertheless, in the region where strong gradient exists,
around~$r=6$, the denominator of Eq.~(\ref{eq:Chargedensity}) remains
close to unity. Indeed, in the differentially rotating part, we have
$r\,\Omega(r)/v_{\rm max} \lesssim 1/2$ implying
$(1-r^2\,\Omega^2/c^2)^{-1} \lesssim 4/3$, therefore the
non-relativistic calculations are still valid in this part of the disk
with an error less than roughly 20\%. Actually, the exact rotation
profile is not important in our study, we just want to demonstrate
that the diocotron instability still exists in the relativistic
regime.

We generalise the study presented in \cite{2007A&A...464..135P} by
including the relativistic effects, i.e. relativistic speed of the
flow and electromagnetic field perturbation.

We start with a non-relativistic rotation profile such that $\Omega_*
\, R_2 \ll c$ and slowly increase $R_2$ (as well as $W_1, W_2, R_1$ to
maintain their ratio constant) in order to approach the speed of light
for the maximal rotation rate of the plasma column.

First, we computed the growth rates when an outer wall is presented.
Second, we remove this wall, imposing outgoing wave solutions and
finally compare both situations.

\subsubsection{Outer wall}

\begin{figure}[htbp]
  \centering 
  \begin{tabular}{cc}
    \includegraphics[scale=0.6]{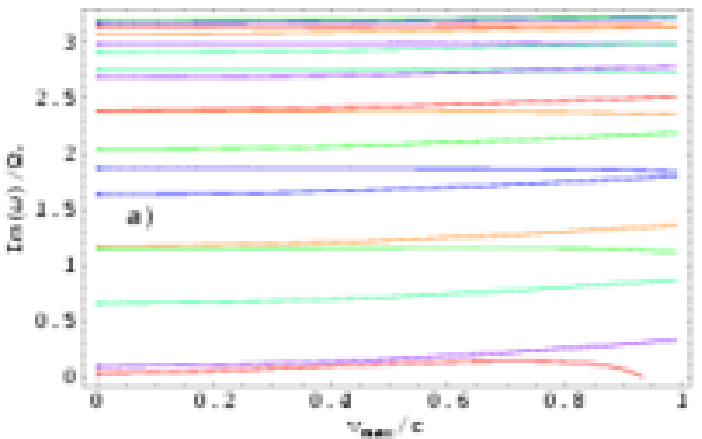} & 
    \includegraphics[scale=0.6]{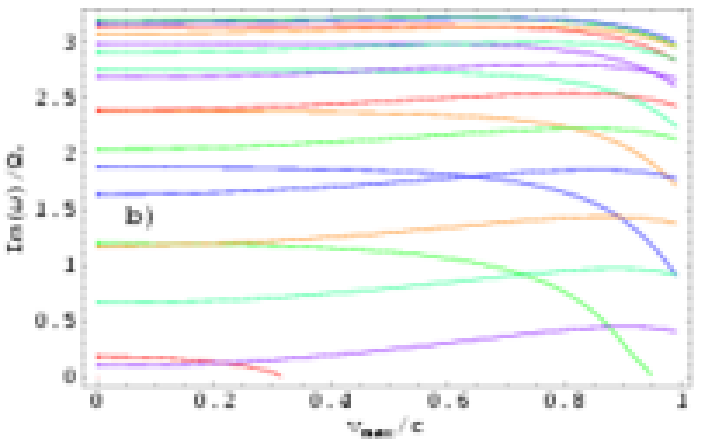} 
  \end{tabular}
  \caption{Evolution of the growth rates~${\rm Im}(\omega)$ 
    of the profiles $\Omega_1$ for increasing maximal speed of the
    column. The column is extending from $R_1=1$ to $R_2=20$ in
    Fig.~a), and from $R_1=1$ to $R_2=10$ in Fig.~b). Each coloured
    curve depicts a different azimuthal mode number~$l$, they are not
    labeled. The outer wall is located at $R_2$.}
  \label{fig:Elec1}
\end{figure}

To remain fully self-consistent, we only consider an uniform applied
external magnetic field.  The maximum growth rates, normalised to the
angular velocity of the neutron star, for the first rotation curve,
$\Omega_1$, for each mode~$l$, is shown in Fig.~\ref{fig:Elec1}a) for
an extension from $R_1=1$ to $R_2=20$. Each coloured curve depicts a
different azimuthal mode number~$l$. However, they are not labeled
because we want only to demonstrate the stabilisation of all modes.
The precise value of the eigenvalues associated with a particular
mode~$l$ therefore does not matter in our present study.  The most
important characteristic is the behaviour of the full set of unstable
eigenvalues when the system becomes relativistic.

The profile having the steepest gradient possesses the largest number
of excited unstable modes because it corresponds to the case where the
smallest scales appear, i.e. $l$ large. In the non-relativistic limit,
$v_{\rm max} \ll c$, the largest growth rate, for $l = 8$ has a value
of $\gamma_{\rm max} = 3.2$. The spectrum is exactly the same as in
\cite{2007A&A...464..135P}. When the maximal speed is increased, the
growth rates vary significantly but we do not observe the
stabilisation effect except for the mode $l=1$ which disappears for
$v_{\rm max}=0.936$.  For relativistic speed, $v_{\rm max}\la c$, the
other modes keep roughly their growth rate at their value for the
non-relativistic instability. Several of them even increase.

In order to demonstrate the tendency towards stabilisation in the
electrosphere, we reduce the size of the plasma column. For instance,
we choose the plasma extension from $R_1=1$ to $R_2=10$.  The new
growth rate are shown in Fig.~\ref{fig:Elec1}b). The diocotron
instability now tends to stabilise for all modes~$l$. Close to the
speed of light, the growth rates start to decrease significantly. Note
that the $l=1,2$ modes already disappeared, Fig.~\ref{fig:Elec1}b).

\begin{figure}[htbp]
  \centering 
  \begin{tabular}{cc}
    \includegraphics[scale=0.6]{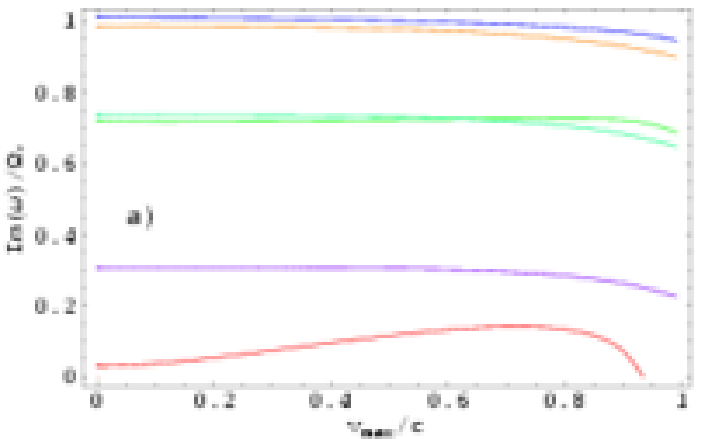} & 
    \includegraphics[scale=0.6]{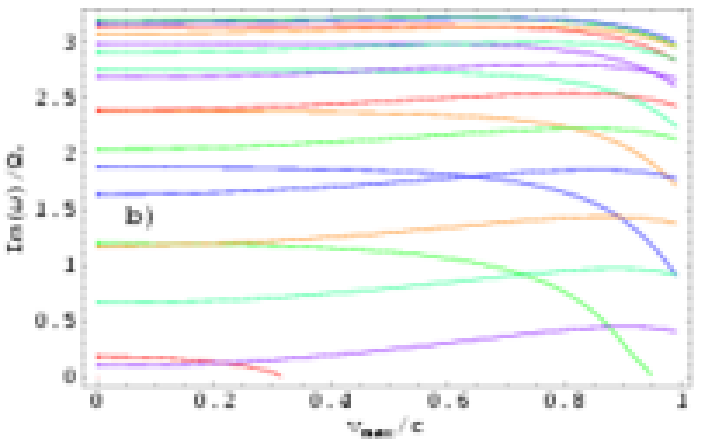}
  \end{tabular}
  \caption{Evolution of the growth rates~${\rm Im}(\omega)$ 
    of the profiles $\Omega_2$ for increasing maximal speed of the
    column. The column is extending from $R_1=1$ to $R_2=20$ in
    Fig.~a), and from $R_1=1$ to $R_2=10$ in Fig.~b). The outer wall
    is located at $R_2$.}
  \label{fig:Elec2}
\end{figure}

The second steepest profile possesses less unstable modes as we would
expect due to the fact that only larger scale structures can emerge
with this slope of the differential rotation, Fig.~\ref{fig:Elec2}a).
Here, the tendency to stabilise the diocotron instability is more
evident. The mode $l=1$ disappears as in the previous case. The other
modes commence to show a significant decrease in their growth rate
when approaching the ultra-relativistic limit.  Here again, we checked
that for low speeds, we recover the non-relativistic spectrum with
good accuracy. For the narrower layer, the stabilisation is fully
achieved for all the modes, Fig.~\ref{fig:Elec2}b).

\begin{figure}[htbp]
  \centering 
  \begin{tabular}{cc}
    \includegraphics[scale=0.6]{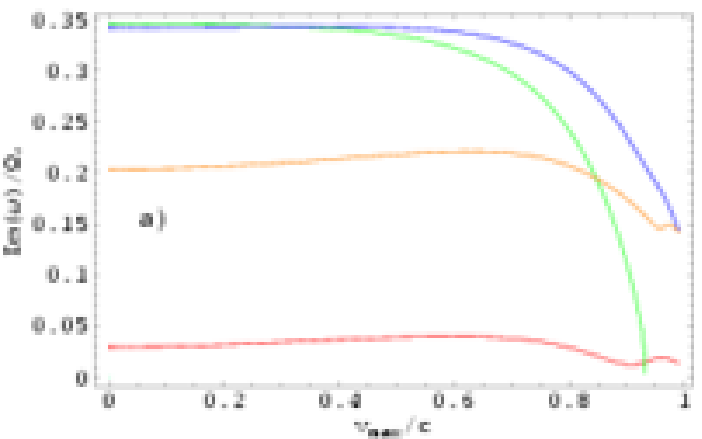} &
    \includegraphics[scale=0.6]{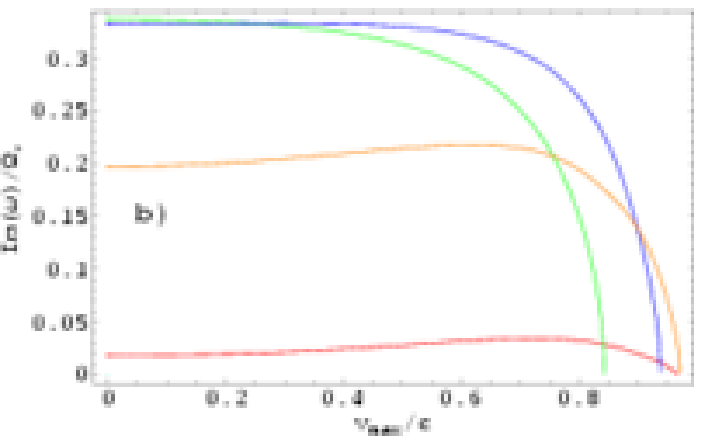} 
  \end{tabular}
  \caption{Evolution of the growth rates~${\rm Im}(\omega)$ 
    of the profiles $\Omega_3$ for increasing maximal speed of the
    column. The column is extending from $R_1=1$ to $R_2=20$ in
    Fig.~a), and from $R_1=1$ to $R_2=15$ in Fig.~b).  The outer wall
    is located at $R_2$.}
  \label{fig:Elec3}
\end{figure}

Finally, the third smooth profile has only four unstable modes,
Fig.~\ref{fig:Elec3}a). The stabilisation effects are clearly seen for
the modes $l=2$ whereas the other modes are near to full
stabilisation. Reducing the size of the electrosphere, here again we
observe full disappearance of the diocotron instability,
Fig.~\ref{fig:Elec3}b).

The proof of the relativistic stabilisation effect are therefore
obvious for the profiles $\Omega_2$ and $\Omega_3$. It is clearly seen
that the diocotron instability is suppressed when the maximal speed
approaches the speed of light with a steep decrease in growth rate for
$v_{\rm max} \la c$.  These examples undoubtfully reveal the influence
of relativistic and electromagnetic effects towards stabilisation of
the non-neutral plasma in the pulsar electrosphere.

To understand the change in the behaviour of the diocotron instability
when simply changing the size of the plasma column, or equivalently,
the dimension of the electrosphere, we need to understand how this
instability is generated.  The instability is launched from the
resonance between the motion of the wave pattern related to the
perturbation and particles in the flow. It is readily seen from the
denominator of Eq.~(\ref{eq:ValPropDiocRelat}). The corotation
radius~$r_{\rm c}$ satisfies
\begin{equation}
  \label{eq:RCorot}
  Re(\omega) = l \, \Omega(r_{\rm c})
\end{equation}
In order for the relativistic effects to play a significant role, the
speed in the vicinity of the corotation region should be close to the
speed of light. If the tail of the rotation curve is long, the outer
edge of the plasma attains speeds close to~$c$ while the speed near
the corotation radius remains non-relativistic. Thus, the growth rates
are not significantly affected by the relativistic effects. However,
reducing the tail of the rotation curve allows the region near
corotation to reach higher velocities. Therefore, the stabilisation
starts to set in and the diocotron instability can be suppressed.

\subsubsection{Outgoing waves}

We performed a second set of calculations by removing the outer wall
assumption and enforce outgoing electromagnetic waves propagation into
vacuum. The boundary condition on the outer plasma/vacuum interface
has been discussed in Sect.~\ref{sec:Boundary}.
  
\begin{figure}[htbp]
  \centering 
  \begin{tabular}{cc}
    \includegraphics[scale=0.6]{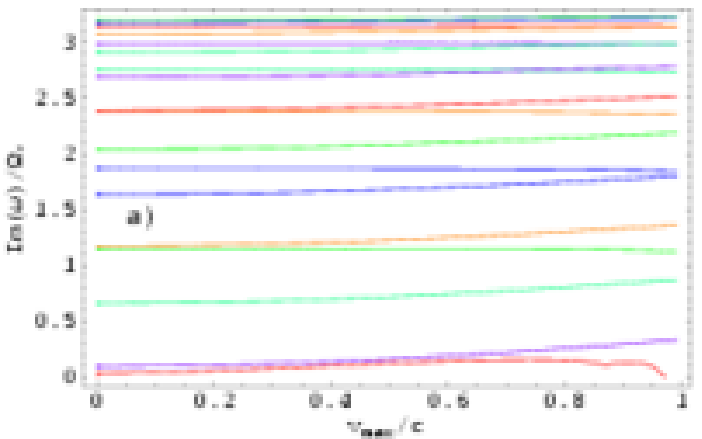} & 
    \includegraphics[scale=0.6]{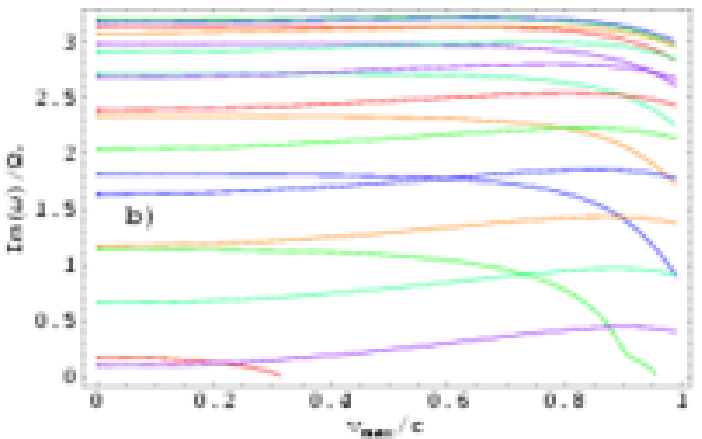} 
  \end{tabular}
  \caption{Same as Fig.~(\ref{fig:Elec1}) but 
    outgoing waves boundary conditions are applied.}
  \label{fig:ElecWave1}
\end{figure}

\begin{figure}[htbp]
  \centering
  \begin{tabular}{cc}
    \includegraphics[scale=0.6]{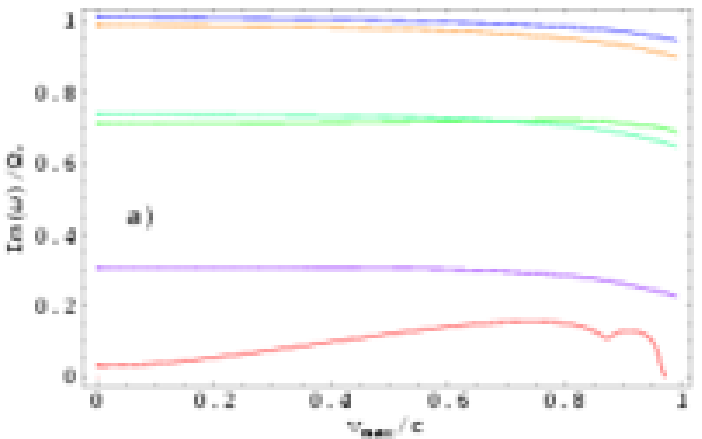} & 
    \includegraphics[scale=0.6]{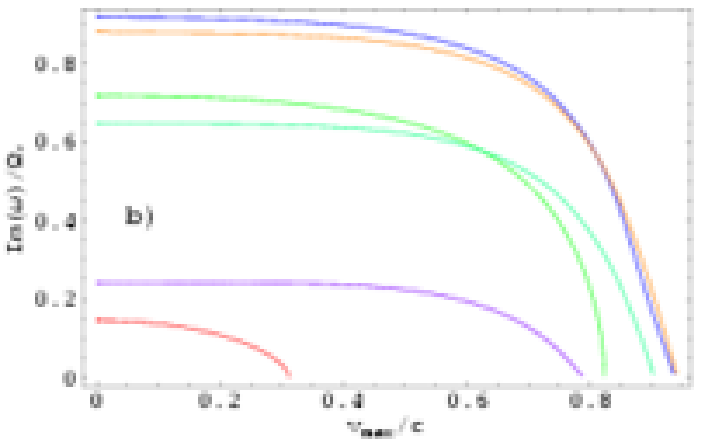}
  \end{tabular}
  \caption{Same as Fig.~(\ref{fig:Elec2}) but 
    outgoing waves boundary conditions are applied.}
  \label{fig:ElecWave2}
\end{figure}

\begin{figure}[htbp]
  \centering 
  \begin{tabular}{cc}
    \includegraphics[scale=0.6]{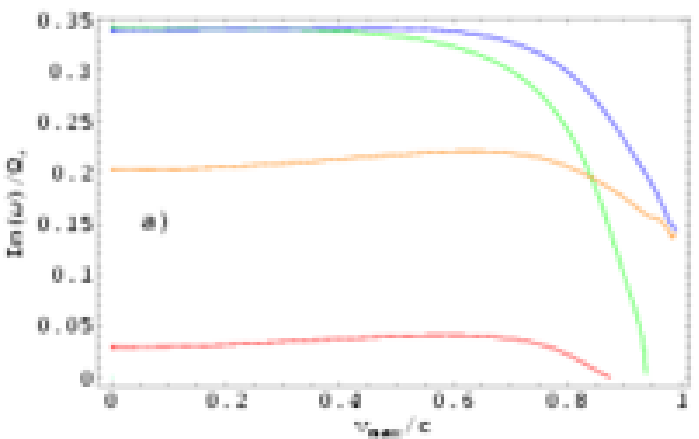} &
    \includegraphics[scale=0.6]{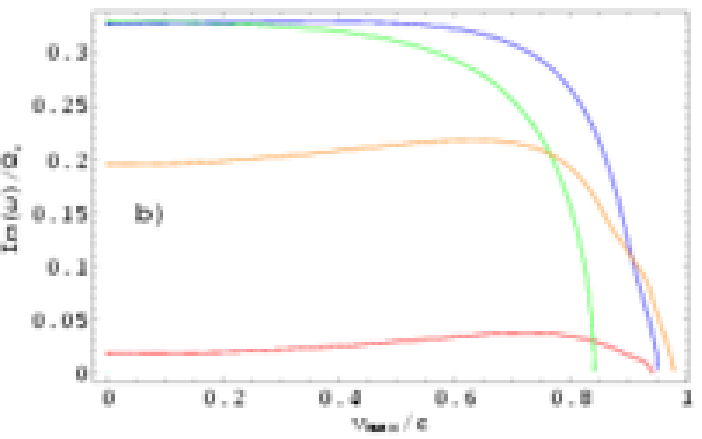} 
  \end{tabular}
  \caption{Same as Fig.~(\ref{fig:Elec3}) but 
    outgoing waves boundary conditions are applied.}
  \label{fig:ElecWave3}
\end{figure}

We use exactly the same configurations and rotation profiles presented
in the previous section. The only change comes from the outer boundary
condition, namely $\phi_{\rm III}(W_2)=0$ which is replaced by
Eq.~(\ref{eq:Jump3}). The results are shown in
Fig.~\ref{fig:ElecWave1}, \ref{fig:ElecWave2} and \ref{fig:ElecWave3}
for the rotation curve $\Omega_1$, $\Omega_2$ and $\Omega_3$
respectively. Comparing both situations, the growth rates are not
significantly affected by wave emission. Note however, that because
energy is carried away by Poynting flux, the instability grows
slowlier than in the previous case.

\begin{figure}[htbp]
  \centering 
  \includegraphics[scale=0.6]{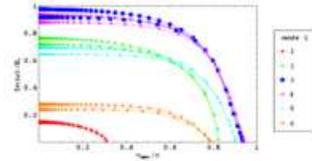} 
  \caption{Comparison of the outer wall and outgoing wave 
    boundary conditions for the profile $\Omega_2$,
    Fig~\ref{fig:Elec2}~b) and Fig~\ref{fig:ElecWave2}~b). The growth
    rates for outgoing waves are almost always slightly smaller.}
  \label{fig:Compar}
\end{figure}

In Fig.~\ref{fig:Compar}, we show a comparison between both boundary
conditions for the profile~$\Omega_2$. It is clearly seen that the
growth rates are relatively insensitive to the nature of the boundary.
Nevertheless, there is a tendency to decrease the growth rates when
outgoing waves are present.

\section{CONCLUSION}
\label{sec:Conclusion}

We developed a numerical code to compute the eigenspectra and
eigenfunctions of the diocotron instability including electromagnetic
and relativistic effects. In the non-relativistic limit, we recovered
and confirmed the results obtained by another technic using
pseudo-spectral method by Chebyshev expansion,
\cite{2007A&A...464..135P}.  Unstable modes are computed for a uniform
external applied magnetic field and arbitrary velocity, density and
electric field profiles.  Application to a plasma column as well as to
the pulsar electrosphere have been shown.  In both cases, the
diocotron regime gives rise to instabilities with decreasing growth
rates when the maximal speed of the flow becomes relativistic.
Whereas the growth rates can be comparable to the rotation period of
the neutron star in the non-relativistic limit, it is found that for
special rotation profiles, the diocotron instability is completely
suppressed in the relativistic regime.  Including electromagnetic wave
emission from the electrospheric plasma does not change drastically
these conclusions.

What therefore happens to the plasma in the vicinity of the light
cylinder needs a more general treatment including inertia of the
particles, because the plasma kinetic energy becomes comparable to the
magnetic field energy density. At the light cylinder, the plasma will
therefore be subject to the magnetron instability. It is the most
general case, including relativistic flow, electromagnetic
perturbation and inertia of the particles.  The study of the magnetron
instability in a pulsar electrosphere is the aim of a forthcoming
paper.

Last but not least, the influence of finite temperature in the plasma
on the diocotron or magnetron instability would require a kinetic
treatment of the stability via the Vlasov-Maxwell equation. This is
also left for future work.

\begin{acknowledgements}
  I am grateful to Jean Heyvaerts and John Kirk for helpful
  suggestions and comments. This work was supported by a grant from
  the G.I.F., the German-Israeli Foundation for Scientific Research
  and Development.
\end{acknowledgements}


\end{document}